\documentclass[twocolumn]{aa}
\usepackage{graphicx}
\usepackage{txfonts}
\newcommand{\eo}{${\mathrm{E_0}}$}
\newcommand{\ep}{${\mathrm{E_p}}$}
\newcommand{\emax}{${\mathrm{E_{max}}}$}

\newcommand{\lxlg}{$S_x / S_{\gamma}$}
\begin{document}
   \title{Spectral analysis of 35 GRBs/XRFs observed with HETE-2/FREGATE}

   \subtitle{}

   \author{C. Barraud,\inst{1}
	  J-F. Olive,\inst{2}
          J.P. Lestrade,\inst{1}\fnmsep\thanks{On leave from Mississippi State University}
	  J-L. Atteia,\inst{1}
	  K. Hurley,\inst{3}
	  G. Ricker,\inst{4}
          D. Q. Lamb,\inst{5}
          N. Kawai,\inst{6,7}
	  M. Boer,\inst{2}
	  J-P. Dezalay,\inst{2}
          G. Pizzichini,\inst{12}
	  R. Vanderspek,\inst{4}         
	  G. Crew,\inst{4}         
	  J. Doty,\inst{4}         
	  G. Monnelly,\inst{4}         
	  J. Villasenor,\inst{4}         
	  N. Butler,\inst{4}         
	  A. Levine,\inst{4}         
          A. Yoshida,\inst{7,8}
          Y. Shirasaki,\inst{10}
          T. Sakamoto,\inst{6,7,11}
          T. Tamagawa,\inst{7}
          K. Torii,\inst{7}
          M.~Matsuoka,\inst{9}
          E.~E.~Fenimore,\inst{11}
          M.~Galassi,\inst{11}
          T.~Tavenner,\inst{11}
          T.~Q.~Donaghy\inst{5}
          C.~Graziani\inst{5}
          \and
          J.G. Jernigan\inst{3}
         }

   \offprints{C. Barraud}

   \institute{Laboratoire d'Astrophysique, Observatoire Midi-Pyr\'en\'ees, 31400 Toulouse, France 
              \email{barraud@ast.obs-mip.fr}
	      \and
	      C.E.S.R., Observatoire Midi-Pyr\'en\'ees, 31028 Toulouse Cedex, France 
              \and
              UC Berkeley Space Sciences Laboratory, Berkeley CA 94720-7450, USA 
              \and
	      Center for Space Reserach, MIT, Cambridge, MA , USA 
	      \and
              Department of Astronomy and Astrophysics, University of Chicago, 5640 South Ellis Avenue, Chicago, IL 60637, USA 
              \and
	      Department of Physics, Tokyo Institute of Technology, 2-12-1 Ookayama, Meguro-ku, Tokyo 152-8551, Japan. 
	      \and
	      RIKEN (Institute of Physical and Chemical Research), 2-1 Hirosawa, Wako, Saitama 351-0198, Japan 
              \and
              Department of Physics, Aoyama Gakuin University, Chitosedai 6-16-1 Setagaya-ku, Tokyo 157-8572, Japan 
              \and
              Tsukuba Space Center, National Space Development Agency of Japan, Tsukuba, Ibaraki, 305-8505, Japan 
              \and
              National Astronomical Observatory, Osawa 2-21-1, Mitaka,  Tokyo 181-8588 Japan 
              \and
              Los Alamos National Laboratory, P.O. Box 1663, Los Alamos, NM, 87545 
              \and
              Consiglio Nazionale delle Ricerche (IASF), via Piero Gobetti, 101-40129 Bologna, Italy 
             }

   \titlerunning{Spectral analysis of 35 GRBs}
   \authorrunning{C. Barraud et al.}

   \date{Received ; accepted }

   \abstract{We present a spectral analysis of 35 GRBs detected with the HETE-2 gamma-ray
detectors (the FREGATE instrument) in the energy range 7-400 keV.
The GRB sample analyzed is made of GRBs localized with the Wide Field X-ray Monitor 
onboard HETE-2 or with the GRB Interplanetary Network.
We derive the spectral parameters of the time-integrated spectra,
and present the distribution of the low-energy photon index, alpha, and
of the peak energy, \ep . We then discuss the existence and nature of the recently
discovered X-Ray Flashes and their relationship with classical GRBs.
   }

   \maketitle
%

\section{Introduction} \label{introduction}

The radiation mechanisms at work during the prompt phase of GRBs
remain poorly understood, despite the observation of hundreds
of GRB spectra and extensive theoretical work 
(e.g. \cite{cohe97,daig98,lloy00,mesz00,pana00,pira00,zhan02}).
One of the reasons for this situation is the lack of broad-band coverage
of this brief phase of GRB emission (contrary to the afterglows
which can be observed from hours to days after the burst).
Recently, however, several instruments have extended
the spectral coverage of the prompt GRB emission to the X-ray range, 
and to optical wavelengths in the case of GRB990123 (\cite{aker99}), raising hopes
for a better understanding of this crucial phase of GRB emission.

We present here the broad-band spectra of 35 GRBs observed
by HETE-2/FREGATE in the energy range 7-400 keV.
We analyse the time-integrated spectra in order to derive
the distribution of their peak energies and of their low-energy spectral
indices.
We also discuss the existence of a possible
new class of soft bursts, called X-ray flashes.

HETE-2's unique instrument suite provides broadband energy coverage
of the prompt emission extending into the X-ray range. The three
instruments include a gamma-ray spectrometer
sensitive in the range 7-400 keV (FREGATE, \cite{atte02}), a Wide Field
X-ray Monitor sensitive in the range 2-25 keV (WXM, \cite{kawa02}) and
a CCD based Soft X-ray Camera working in the range 1-14 keV (SXC, \cite{vill02}).
In this paper we restrict our analysis to FREGATE data
because this instrument, with its larger field of view, detects 
about two times more GRBs than WXM (the Half Width at Zero Maximum is 70$^\circ$ 
for FREGATE compared to 40$^\circ$ for the WXM) and because in most cases FREGATE data
are sufficient to determine the GRB spectral parameters.
We finally note that FREGATE offers for the first time a
continuous coverage from 7 keV to 400 hundred keV
{\it with a single instrument}.
This eliminates any possible problems caused by normalizing
the responses of different instruments to one another.
This characteristic of FREGATE appears
essential when we try to understand whether events seen at low energies
are of the same nature as classical GRBs seen at higher energies.

This work follows many previous studies which contributed to
our understanding of the GRB spectral properties at gamma-ray
energies (\cite{band93,pree98}) and in hard X-rays (\cite{stro98,fron00a,kipp01}).
This paper is the first of a series devoted to the spectral analysis of
the GRBs detected with HETE-2. Forthcoming papers will discuss the spectral
{\it evolution} of bright GRBs and the broad-band spectral distribution from
2 keV to 400 keV by combining
the data from FREGATE and the WXM for the events which are detected by both
instruments.


\section{Spectral calibration} \label{calibration}

FREGATE has been designed to provide reliable spectral data on the
gamma-ray bursts and a detailed description of the instrument can be found 
in Atteia et al. (2002).
The detectors use cleaved NaI crystals encapsulated in a beryllium
housing offering a good sensitivity at low energies (the transmission
of the window is greater than 65\% at 6 keV).
A graded shield made of lead, tantalum, tin, copper and aluminum reduces
the background and eliminates many GRBs arriving at angles more than 70$^\circ$ off-axis.
Two on-board sources of Baryum 133 provide a continuous monitoring
of the gain of the 4 detectors. The gain adjustment is done on the ground;
there is no automatic gain control.
The whole instrument has been carefully simulated with GEANT
software from the CERN (see http://wwwinfo.cern.ch/asd/geant/index.html) and the
output of the simulation program has been checked and validated against
extensive calibrations done with radioactive sources (9 sources,
11 energies at 5 angles).
Finally the in-flight spectral response has been checked with the
Crab nebula as described in \cite{oliv02a} and briefly explained below.

As a consequence of the antisolar orientation of HETE-2, the Earth occults 
the sources in the field of view of FREGATE once per orbit.
We used this feature, to reconstruct the spectrum of the Crab nebula 
from the size of the Crab occultation steps measured at various energies. 
We then used the standard spectral analysis of FREGATE to derive the spectral
parameters of the Crab nebula, assuming a power law spectrum. 
\cite{oliv02a} find a spectral index of $2.16 \pm 0.03$ in the range 6-200 keV, 
and a normalization of 7.2 10$^{-3}$ ph cm$^{-2}$ s$^{-1}$ keV$^{-1}$ at 30 keV, 
fully consistent with values measured by other instruments at these energies.

\section{Spectral analysis} \label{analysis}

\begin{figure}
\resizebox{\hsize}{!}{\includegraphics{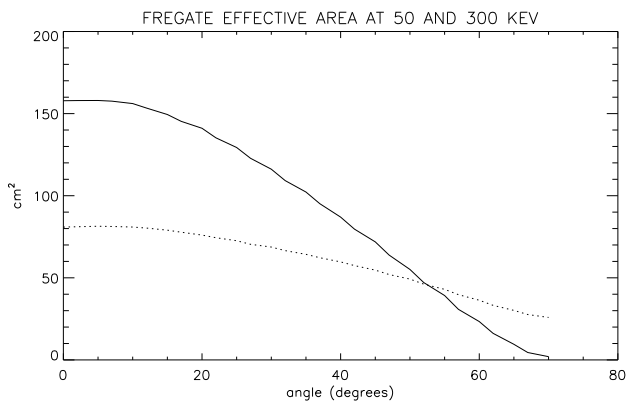}}
\caption{FREGATE effective area as a function of the burst angle
at 50 keV (solid line) and 300 keV (dotted line).}
\label{effarea}
\end{figure}

Figure \ref{effarea} displays the angular response of FREGATE.
This figure emphasizes the importance of knowing the GRB off-axis angle
to perform a reliable spectral analysis.
Our spectral analysis includes the following steps:

 \begin{enumerate}
      \item Selection of the burst sample (section \ref{sample}).
      \item Construction of gain corrected spectra and addition of the spectra from the 4 detectors.
      \item Determination of the maximum energy \emax\ at which there is still
            some signal from the burst (Sect. \ref{fits}).
      \item Determination of the arrival angle of the burst to within 5 degrees 
      \item Spectral deconvolution with XSPEC and determination of the burst fluence and spectral parameters.
\end{enumerate}

\subsection{The GRB sample} \label{sample}

From October 2000 to mid November 2002, FREGATE detected
88 confirmed GRBs of which 61 were within the 70$^\circ$ field of view
(FOV) of the detector.\footnote{FREGATE continuously records
the count rates in 4 energy ranges ([6-40],
[6-80], [32-400], and [$>$400] keV,  see \cite{atte02}).
GRBs outside the FOV have almost no counts below 80 keV, providing
a reliable way to recognize the GRBs which are within the FOV of FREGATE,
even in the absence of a localization.}
However not all these 61 bursts were localized and since we cannot
perform accurate spectral studies of GRBs with unknown off-axis angles,
we concentrate here on the analysis of the spectra of 35 GRBs which
have been localized with the WXM or the Interplanetary Network (IPN).
The list of these 35 GRBs is given in Table 1 with their off-axis angle
and their duration T$_{90}$.
Thirty three GRBs are within 60$^\circ$ of the detector axis.
Figure \ref{effarea} shows that for these bursts we have an efficiency $\geq 15\%$
at low (E = 50 keV) energies and the instrument response remains constant for small variations in angle.
Beyond 60$^\circ$, the efficiency of the detector is very low at low energies
and its energy response changes quickly for small changes in angle. 
As a consequence, the parameters for GRB020214 and GRB020418 (at 66$^\circ$ and 64$^\circ$
respectively) must be taken with caution.

Our sample includes two short/hard GRBs: GRB020113 and GRB020531 (\cite{lamb02}) .
In fact only 2 out of the 61 confirmed GRBs within the
FOV of HETE-2/FREGATE are short/hard GRBs (less than 5\%), whereas
about 20\% of the BATSE bursts were short, hard GRBs (\cite{paci99}).  
This is most likely due to two factors:

(1) BATSE was less sensitive to X-ray-rich GRBs (all of which are
long GRBs).  However, a substantial percentage of the GRBs in
the FREGATE sample are X-ray-rich [for example, 9 of the 35 GRBs in
Table 1 have \lxlg $\ge$ 0.6; i.e., nearly 30\% of our sample are X-ray-rich]. 
Therefore the number of GRBs in the FREGATE sample which should
be compared to the BATSE sample is significantly smaller than 61.

(2) Because short, hard GRBs are generally quite hard (by definition),
it is possible that we have classified a few of them as having occurred
outside the FREGATE FOV when in fact they occurred within the FREGATE
FOV (see footnote 1).

Factor (1) could reduce the size of the sample of GRBs that should be
compared with the BATSE result by as much as 30\% (0.7 x 61 = 42).  This
would make the statistics to $\sim$2/42, which corresponds to $\sim$5\% (with
large errors).  If we have mis-classified 2 or 3 short, hard GRBs as
having occurred outside the FREGATE FOV when in reality they occurred
within the FREGATE FOV, the fraction becomes $\sim$4/42 or $\sim$5/42, which
corresponds to 11\% - 14\% (again with large errors) and which would be
closer to the percentage found by BATSE.

If we consider only those GRBs in the sample of 35 GRBs which have
been localized by the IPN and therefore which form an unbiased sample
with regard to factors (1) and (2) above, the fraction of short, hard
GRBs in this sample is 2/13, which corresponds to 15$^{(+11)}_{(-8)}$ \%
(1-sigma).  This result is fully consistent with the BATSE results.

Consider now the GRBs in Table 1 which have been localized by the WXM.
These bursts are affected by factor (1) but not by factor (2) (by
definition, if the burst has been localized by the WXM, it occurred
within the FREGATE FOV).  However, these bursts are also affected by a
third factor:

(3) Since short, hard GRBs are quite hard (by definition), some have
relatively little emission in the WXM energy band.  Consequently, they
may not be localized, even though they occurred within the WXM FOV (and
therefore within the FREGATE FOV).  GRB 020113 in Table 1 is an example
of such a burst.

That factors (1) and (3) are likely important can be demonstrated by
examining the GRBs in Table 1 which were localized using the WXM.
There are 26 of these, of which only 1 is a short, hard GRB.  The
probability of this happening by chance, assuming that short, hard GRBs
are 20\% of all GRBs, is $2.2 x 10^{-2}$.

\subsection{spectral fits} \label{fits}

GRB photon spectra can in general be fit by the
Band function (\cite{band93}); which is two smoothly connected power laws.
A typical GRB photon spectrum is thus described as follows:

\begin{displaymath}
N(E) = A E^\alpha \exp ({-E \over E_0}) ~~~~~~\mathrm{for}~~~ E \le (\alpha - \beta) E_0 , ~~~~~~\mathrm{and}~~~
\end{displaymath}

\begin{equation}
N(E) = B E^\beta   ~~~~~~\mathrm{for}~~~ E \geq (\alpha - \beta) E_0 , 
\end{equation}

\begin{displaymath}
\mathrm{where}~~~   B = A [(\alpha - \beta) \times E_0]^{(\alpha - \beta)} \times \exp (\beta - \alpha)
\end{displaymath}

Here $\alpha$ is the photon index of the low energy power law, $\beta$
is the photon index of the high energy power law, and $E_0$ is
the break energy.
With this parametrization, the peak energy of the $\nu f_{\nu}$ spectrum
is $E_{p} = E_0 \times (2+\alpha)$.
\ep\ is well defined for $\alpha \ge -2$ and $\beta < -2$.

The first step of our processing was the definition of the
spectral range appropriate for the fitting procedure.
On the low energy side the limit is set to 7 keV for instrumental
reasons (the electronics threshold is not well modeled below 7 keV).
On the high energy side, we computed the energy \emax\ such that
the signal in the range [\emax\ - 400] keV was only
two sigmas above background.
The spectral fit is done in the range [7 - \emax ] keV,
\emax\ is given in Table \ref{grblist}.

Most of the time this energy range is not broad enough to allow
the unambiguous determination of the 4 parameters of the Band function.
Consequently, we decided to fit the observed spectra using only the
low energy part of the Band function and the spectral break.
The definition of \ep\ is not affected by the choice of this model.
In the following, this model is called the cutoff power law model,
and it is defined as

$$N(E) = A E^\alpha \exp ({-E \over E_0}) $$

The spectral deconvolution is done with XSPEC, using
gain corrected spectra (allowing us to co-add the 4 detectors) and
response matrices constructed from a detailed Monte Carlo simulation of
the instrument (\cite{oliv02b}).
We include systematic errors at a level of 2\% in the fitting procedure.
This level of systematic errors is required to correctly fit a very bright
burst from SGR 1900+14 with a smooth spectral shape. 
 
The parameters resulting from the fit and their 90\% errors are given
in Table \ref{grblist}. The cutoff power law model
provides a good fit to our data as seen by reduced $\chi ^2$ values close to unity
(see however section \ref{peculiar} below).
Examples of spectral fits are given in Fig.~\ref{twospectra}.

It is noted that the cutoff energy E$_0$ is well constrained
for only 19 GRBs (out of 35).
These events are identified in Table \ref{grblist} by their names in boldface.
In the following we call these GRBs group A, and group B the 16 GRBs
whose \eo\ is not constrained.
Group B bursts are either faint bursts with not enough counts at high energy 
to constrain \eo\, or GRBs which have their \eo\ outside the energy range 
of FREGATE.
It is tempting to fit the spectra of group B bursts with a simple
power law (since we have no constraint on the cutoff energy) which would
give stronger constraints on $\alpha$.
It is not a good idea, however, to use a different model for one part of the sample
because this introduces a bias in the measured values of the spectral parameters
(see also \cite{band93}).
We thus keep the cutoff power law model, even when we have no constraint on \eo , 
and indicate in Table \ref{grblist} the best fit parameters and their errors
for this model. We now discuss the distribution of the spectral
parameters obtained with this procedure.

\begin{figure*}
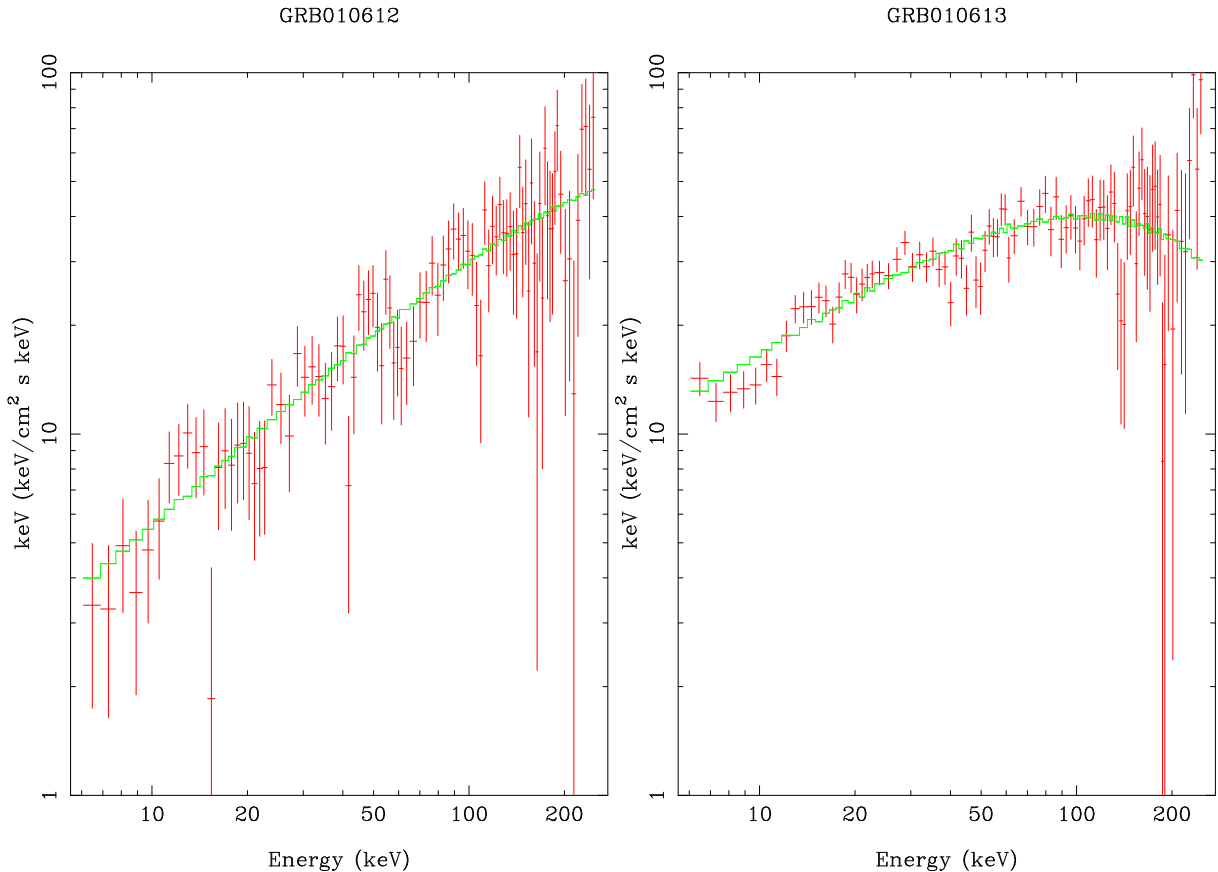

\includegraphics[width=8cm]{GRB010612papier.ps}
\includegraphics[width=8cm]{GRB010613papier.ps}
\caption{Spectral fits of GRB010612 and GRB010613 with a power law with a cut-off
at 592 keV and 176 keV respectively.}
\label{twospectra}
\end{figure*}

\subsection{Peculiar bursts} \label{peculiar}
We discuss here a few bursts which required special care in their spectral analysis.

\subsubsection{GRB020113 and GRB020531} 
As mentioned in section \ref{sample}, GRB020113 and GRB020531 are Short/Hard GRBs. 
We provide their spectral parameters in Table \ref{grblist}, but we do not consider 
them further in the rest of the analysis.

\subsubsection{GRB001225 and GRB020813}
GRB001225 and GRB020813 are the two strongest GRBs detected by FREGATE.  
The spectra of these two bursts are adequately fit by a cutoff power law, 
except for a deficit of photons between 30 and 55 keV
(see Appendix 1), which is responsible for
the high $\chi^2$ values given in Table \ref{grblist}. When we exclude the
region 30-55 keV for the spectral analysis we obtain a reduced $\chi ^2$
of 1.24 for 96 DOF for GRB001225 and a reduced $\chi ^2$ of 1.20 for 97 DOF
for GRB020813. The origin of this feature in the spectra of these two bright bursts
is being investigated with a great care due to its proximity with the k-edge of iodine.
The exclusion of this region doesn't change the global parameters
for these two spectra.

\subsubsection{GRB021104}
This burst cannot be fit with a single power law and requires a break. The break energy
is too close to the low energy threshold of FREGATE to allow a reliable estimate
of the low energy spectral index $\alpha$. We thus decided to freeze $\alpha$ to $-1.0$
in the spectral fit.

\begin{table*}
\caption{GRB list. This table gives the best fit parameters, the reduced
$\chi ^2$ of the fit and the error bars on the spectral parameters.
The names in bold indicate the events of group A (see text).}
\label{grblist}
$$
\begin{array}{rrrrrrrrcrcrrrr}
\hline
\noalign{\smallskip}
\mathrm{Name} & 
\mathrm{Time} & 
\mathrm{Duration}^{\mathrm{a}} & 
\mathrm{Angle} & 
\mathrm{Loc.}^{\mathrm{b}} & 
\mathrm{E}_{max} & 
{\chi}^2 \mathrm{/ DOF}^{\mathrm{c}} & 
{- \alpha} & 
\mathrm{errors}& 
\mathrm{E}_0 & 
\mathrm{errors} & 
\mathrm{E}_p& 
\mathrm{S}_x ^{\mathrm{d}} & 
\mathrm{S}_{\gamma} ^{\mathrm{e}} & 
\mathrm{S}_x / \mathrm{S}_{\gamma} \\
     & 
\mathrm{SOD}  & 
\mathrm{sec}  & 
\mathrm{deg.} &
     & 
\mathrm{keV}  &
     &
     &
     &
\mathrm{keV}  &
     &
     &
     &
     &
     \\
\noalign{\smallskip}
\hline 
\noalign{\smallskip}
 {\bf 001225} & 25759 & 32.3 & 37 & \mathrm{I}   & 400 & 1.78~/112 & 1.17 & [1.16;1.18] &   283 & [271-296] &  235 & 190 & 1140 &  .17 \\
 {\bf 010126} & 33162 &  7.7 & 50 & \mathrm{I}   & 220 & 1.11~/84  & 1.06 & [0.82;1.26] &   115 & [ 72-218] &  108 & 7.7 & 29.9 &  .26 \\
       010213 & 45332 & 20.9 & 14 & \mathrm{W}   & 250 & 1.11~/89  & 2.14 & [1.83;2.55] & 10000 & [370-10^4] &  <20 & 1.8 &  2.4 &  .75 \\
       010225 & 60733 &  7.2 & 23 & \mathrm{W}   &  70 & 1.16~/42  & .89  & [-1.8;2.14] &    22 & [  5-10^4] &   24 & 1.1 & 0.66 &  1.7 \\
{\bf 010326A} & 11701 & 23.0 & 60 & \mathrm{I}   & 250 &  .67~/89  & .894 & [0.66;1.09] &   260 & [167-484] &  287 &  16 &  160 &  .10 \\
      010326B & 30792 &  3.2 & 17 & \mathrm{W}   & 120 &  .97~/58  & 1.12 & [0.31;1.71] &    69 & [ 25-10^4] &   61 & 1.5 &  3.1 &  .48 \\
       010612 &  9194 & 74.1 & 14 & \mathrm{W}   & 250 &  .85~/89  & 1.22 & [1.07;1.31] &   592 & [274-10^4] &  462 & 6.8 &   49 &  .14 \\
 {\bf 010613} & 27235 & 152. & 36 & \mathrm{W}   & 250 & 1.29~/89  & 1.40 & [1.33;1.47] &   176 & [139-235] &  106 &  70 &  203 &  .34 \\
 {\bf 010629} & 44468 & 15.1 & 28 & \mathrm{W/I} & 200 &  .91~/81  & 1.17 & [1.03;1.31] &    59 & [ 48-75 ] &   49 &  16 &   26 &  .62 \\
 {\bf 010921} & 18950 & 24.6 & 45 & \mathrm{W/I} & 200 & 1.20~/81  & 1.49 & [1.43;1.56] &   206 & [158-287] &  105 &  38 &  102 &  .37 \\
       010923 & 33870 &  3.8 & 58 & \mathrm{I}   & 250 & 1.06~/89  & 1.74 & [1.49;2.04] & 10000 & [347-10^4] & 2600 &  10 &   30 &  .33 \\
 {\bf 010928} & 60826 & 48.3 & 31 & \mathrm{W}   & 400 & 1.17~/112 & .623 &  [.561;.68] &   260 & [220-315] &  358 &  11 &  210 &  .05 \\
       011019 & 31370 & 25.4 & 25 & \mathrm{W}   &  80 &  .76~/45  & 1.75 & [0.04;2.47] &    87 & [ 10-10^4] &   22 & 1.7 &  1.7 &  1.0 \\
       011130 & 22775 & 83.2 & 26 & \mathrm{W}   &  70 &  .97~/42  & 1.08 & [-0.5;2.51] &    32 & [ 16-10^4] &   29 & .79 & 0.68 & 1.16 \\
       011212 & 14642 & 84.4 & 10 & \mathrm{W}   & 150 &  .53~/67  & 1.28 & [-3.0;2.24] &    34 & [  3-10^4] &   25 & 2.3 &  1.7 & 1.35 \\
       011216 & 10524 & 31.8 & 47 & \mathrm{I}   & 100 & 1.33~/52  & 1.82 & [1.54;2.11] & 10000 & [420-10^4] & 1800 & 4.7 & 12.0 &  .39 \\
 {\bf 020113} &  7452 & 1.31 & 34 & \mathrm{I}   & 400 & 1.15~/112 & 0.46 & [0.05;0.78] &   239 & [126-666] &  368 & .54 & 13.3 &  .04 \\
 {\bf 020124} & 38475 & 78.6 & 33 & \mathrm{W}   & 250 &  .91~/89  & 1.10 & [0.98;1.21] &   133 & [101-186] &  120 &  17 &   68 &  .25 \\
 {\bf 020127} & 75444 &  9.3 & 22 & \mathrm{W}   & 200 &  .92~/81  & 1.19 & [1.00;1.36] &   156 & [ 97-330] &  126 & 2.3 &  9.1 &  .25 \\
       020201 & 65828 & 241. & 55 & \mathrm{W}   &  80 &  .95~/45  & 1.67 & [0.71;2.16] &    99 & [ 18-10^4] &   33 &  23 &   28 &  .82 \\
 {\bf 020214} & 67778 & 27.4 & 66 & \mathrm{I}   & 400 & 1.31~/112 & .256 & [.059;.439] &   176 & [145-219] &  307 &  32 &  930 &  .03 \\
 {\bf 020305} & 42925 & 250. & 35 & \mathrm{W}   & 250 & 1.15~/89  & .861 & [.748;.968] &   143 & [113-192] &  163 &  15 &  104 &  .14 \\
       020317 & 65731 &  3.3 & 23 & \mathrm{W}   & 150 & 1.08~/67  & 1.01 & [-.78;1.95] &    44 & [ 11-10^4] &   44 & 1.2 &  1.7 &  .71 \\
 {\bf 020331} & 59548 & 56.5 & 16 & \mathrm{W}   & 400 & 1.09~/112 & .922 & [.812;1.02] &   120 & [ 97-153] &  129 & 8.6 &   45 &  .19 \\
 {\bf 020418} & 63789 & 7.54 & 64 & \mathrm{I}   & 400 &  .92~/112 & 1.10 & [.78;1.37]  &   240 & [150-470] &  216 &  22 &  139 &  .16 \\
       020531 &  1578 & 1.15 & 26 & \mathrm{W/I} & 300 & 1.16~/97  & 1.10 & [0.77;1.30] &   810 & [200-10^4] &  729 & 1.1 & 11.5 &  .10 \\
 {\bf 020801} & 46721 & 336. & 33 & \mathrm{W}   & 300 &  .90~/97  & 1.32 & [1.09;1.53] &   116 & [ 70-252] &   79 &  66 &  163 &  .40 \\
 {\bf 020812} & 38503 & 27.5 & 19 & \mathrm{W}   & 300 &  .98~/97  & 1.03 & [ .72;1.31] &   125 & [ 71-316] &  121 & 5.2 &   23 &  .23 \\
 {\bf 020813} &  9859 &  90. &  4 & \mathrm{W}   & 300 & 1.49~/112 & 1.05 & [1.02;1.07] &   223 & [205-238] &  212 & 152 & 1020 &  .15 \\
 {\bf 020819} & 53855 & 33.6 & 28 & \mathrm{W}   & 300 & 1.06~/97  & 1.03 & [ .93;1.12] &    94 & [ 78-116] &   91 &  16 &   54 &  .30 \\
       021004 & 43573 & 57.7 & 13 & \mathrm{W}   & 300 & 1.11~/97  & 1.64 & [1.32;1.74] &  3000 & [1500-10^4] &1080 & 6.4 &   23 &  .28 \\
       021014 & 23513 & 39.3 & 56 & \mathrm{I}   & 300 & 1.06~/97  & 1.16 & [ .77;1.43] &   504 & [127-10^4] &  423 & 9.2 &   71 &  .13 \\
 {\bf 021016} & 37740 & 81.6 & 36 & \mathrm{W/I} & 230 & 1.12~/86  &  .98 & [ .81;1.13] &   132 & [ 92-211] &  135 &  22 &  113 &  .19 \\
 {\bf 021104} & 25262 & 19.7 & 31 & \mathrm{W}   &  60 & 1.18~/39  &    1 & [-3.0;1.29] &    27 & [ 16-51 ] &   27 & 3.9 &  2.9 & 1.34 \\
       021112 & 12495 &  7.1 & 30 & \mathrm{W}   & 200 & 1.04~/81  & 1.47 & [ .89;1.88] &   186 & [ 48-10^4] &   99 & .93 &  2.5 &  .37 \\

\noalign{\smallskip}
\hline
\end{array}
$$
\begin{list}{}{}
\item[$^{\mathrm{a}}$] Duration T$_{90}$ in the 7-400 keV energy range.
\item[$^{\mathrm{b}}$] This column indicates whether the burst has been localized by the WXM (W) or by the IPN (I).
\item[$^{\mathrm{c}}$] ${\chi}^2$ is the reduced chi square.
\item[$^{\mathrm{d}}$] S$_x$ is the fluence in the energy range 7-30 keV, in units of 10$^{-7}$ erg cm$^{-2}$.
\item[$^{\mathrm{e}}$] S$_{\gamma}$ is the fluence in the energy range 30-400 keV, in units of 10$^{-7}$ erg cm$^{-2}$.
\end{list}
\end{table*}

\section{Discussion} \label{discussion}

\subsection{The distribution of E$_0$ and $\alpha$} \label{distri}

\begin{figure}
\resizebox{\hsize}{!}{\includegraphics{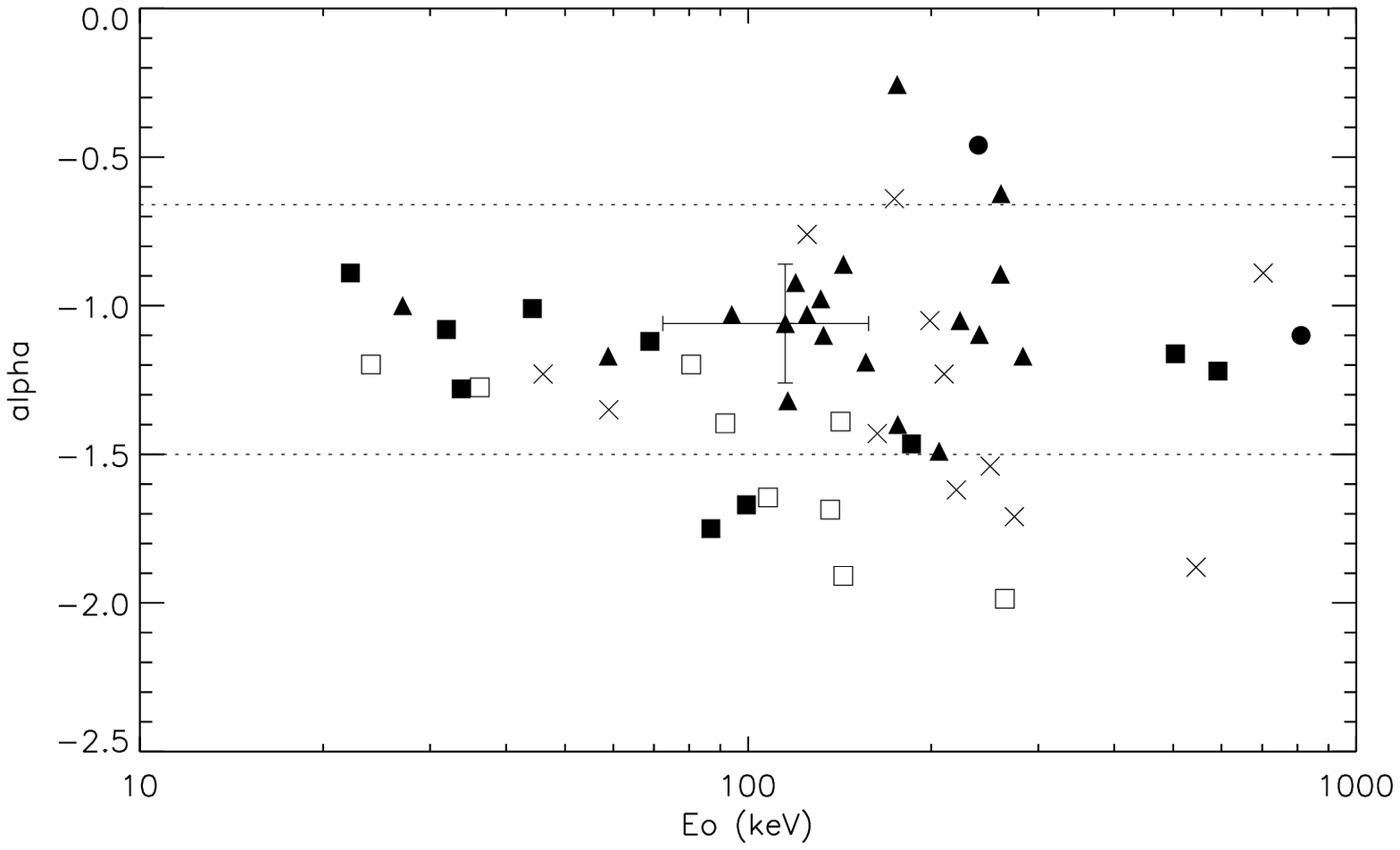}}
\caption{Spectral parameters of 31 GRBs detected by FREGATE (dark symbols). 
As in the following plots group A GRBs (see text) are shown with dark triangles,
group B GRBs with dark squares and the two short/hard GRBs with dark circles.
A typical error bar (on $\alpha$ and E$_0$) is also shown.
Four GRBs with \eo\ above the FREGATE energy range upper limit are not shown in this plot.
For comparison we also display 9 GRBs/XRFs discussed by \cite{kipp01} (empty squares)
and 12 GRBs discussed in \cite{amat02} (crosses).
The dotted horizontal lines delimit the range of $\alpha$ predicted by the synchrotron shock model
(\cite{lloy00}).}
\label{ale0}
\end{figure}

Figure~\ref{ale0} displays $\alpha$ as a function of the cutoff energy E${_0}$
for the 35 GRBs studied here.
We have also plotted for comparison 9 GRBs/XRFs discussed in \cite{kipp01} and 12 GRBs
described in \cite{amat02}.
This figure shows a good agreement between the values measured by BeppoSAX
and FREGATE. We note two features which we will discuss more extensively in the
next sections: most values of $\alpha$ are compatible with the predictions
of the synchrotron shock model and there is a tail of GRBs with \eo\ extending
well below 100 keV.

\subsubsection{The distribution of $\alpha$} \label{alpha}

\begin{figure}
\resizebox{\hsize}{!}{\includegraphics{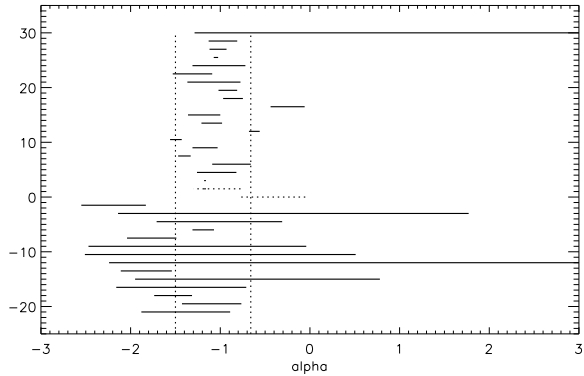}}
\caption{Low energy spectral indices for 35 GRBs detected by FREGATE.
The horizontal lines show the error bars on $\alpha$ for each of the 35 GRBs discussed
in the paper.
GRBs of group A (see text) have a positive ordinate. 
The two short/hard GRBs are indicated with dotted lines.
The two vertical lines indicate the range of $\alpha$ expected for
synchrotron radiation models (e.g. \cite{lloy00}).}
\label{alphafig}
\end{figure}

Table \ref{grblist} and Fig.~\ref{alphafig} show that, with the exception
of GRB010213 ($\alpha = -2.14$, but see discussion below),
and GRB020214 ($\alpha = -.256$), the GRBs
in our sample have $\alpha$ in the range $-3/2$ to $-2/3$, compatible with
the values expected from radiation produced by synchrotron emission
from shock accelerated electrons (\cite{katz94,cohe97,lloy00}).

GRB010213 has the softest spectrum of our sample; its steep spectral index
suggests a GRB having a very low \ep , in any case below 20 keV. 
We consider that the spectral index which is measured
by FREGATE in this case is probably NOT $\alpha$, but rather $\beta$.
In fact a joint fit of the WXM and FREGATE data gives $\alpha = -1.37$, $\beta = -2.14$ and \eo = 4 keV 
(\ep = 2.5 keV) for this burst (\cite{kawa03,saka03}).

GRB020214 has a very hard spectrum, which is definitely not
compatible with synchrotron radiation in its simple form (\cite{lloy02}).
We fit the spectrum of this GRB with the Band function,
in order to check how the fit by a power law with a cutoff affected the value
of $\alpha$. Because this GRB has many high energy photons,
we were able to determine the 4 parameters of the Band function.
We find $\alpha = -0.14 $, \eo\ = 140 keV and
$\beta = -2.11 $ (with the errors [$-0.42$;$+0.12$], [$104$;$213$], [$-10$;$-1.75$], respectively).
The fit with the Band model tends to increase the value of $\alpha$,
and therefore the difference with the canonical synchrotron values.
We should keep in mind however that this GRB arrived on the detectors
with a large off axis angle and that its spectral parameters can change 
quickly if we assume a slightly different angle of incidence (e.g. if we consider an
angle of 68$^\circ$, the 90\% error bar on $\alpha$ is [-0.72;0], marginally
consistent with the predictions of the synchrotron shock model).

\subsubsection{The distribution of \ep} \label{epeak}

In this section we again exclude the two short/hard GRBs.
GRBs of group A have \ep\ which vary from 27 keV to 358 keV,
with 2 events having \ep\ below 50 keV.

GRBs of group B can essentially be divided into 2 subgroups: a few hard bursts which
have \ep\ above the upper energy limit of FREGATE (GRB010612, GRB010923, GRB011216, 
GRB021004, GRB021014) and soft bursts with \ep\ in the range 20-60 keV
(GRB010213, GRB010225, GRB010326B, GRB011019, GRB011130, GRB011212, GRB020201, 
and GRB020317). One event (GRB021112) has \ep\ well within the FREGATE energy
range but is so faint (two times fainter than the faintest 
event of group A) that it can adequately be fit by a simple power law.

Overall we thus have nine GRBs with \ep\ lower than 50 keV in a sample of 33 events
(we define here soft GRBs as GRBs with  \ep\ $\leq$ 50 keV, this is an arbitrary boundary 
but our conclusions do not depend on the choice of this number).

The low \ep\ values for GRBs of group B could be questioned however
because the break energies of these bursts are not well constrained.
We mentioned in section \ref{fits} that these GRBs are fainter events
with less counts than the GRBs of group A. 
We discuss below whether this lack of statistics 
can bias their spectral parameters.
To this aim, we decreased the intensity of GRBs in the group A by a factor
6.5 to construct a new set, A', with the same number of photons as group B
GRBs, and computed the spectral parameters of this new set.
We characterize the spectral hardness of events in groups A, A' and B
by two parameters: an average softness ratio \lxlg\ and the fraction r of events with
\ep\ lower than 50 keV (even if \eo\ is not well constrained for samples
A' and B).
We find that log(\lxlg) = -0.67$\pm$0.08, -0.58$\pm$0.11, and -0.28$\pm$0.09
and  r = 2/19, 2/19 and 7/14 for GRBs of group A, A' and B respectively.
These numbers show that the spectral softness of GRBs in group B is not
an artifact of their smaller number of photons.
We are thus led to the conclusion that group
B contains many intrinsically soft GRBs. 
These soft GRBs have few photons above 50 keV, and for them
the effective energy range of FREGATE is significantly reduced
explaining why their energy spectra can be fit with a single power law.

This analysis also shows that the ratio \lxlg\ is a robust
estimator of the softness of FREGATE GRBs, and in the following
we define soft GRBs as having \ep $\leq$ 50 keV or equivalently \lxlg\ greater than 0.60.

\subsubsection{X-ray rich GRBs} \label{xrich}

The evidence for GRBs with low values of \ep\ (\ep\ $\le$ 50 keV) has been accumulating over
recent years.
In 1998, Strohmayer et al. (1998) studied the X-ray to $\gamma$-ray spectra of 22 GRBs
(they performed joint fits of the data recorded by a proportional counter
and a scintillator spanning energies from 2 to 400 keV).
They found 7 GRBs with \ep\ lower than 10 keV and 5 more with \ep\ lower
than 50 keV, providing the first evidence for a population of soft GRBs.
In the 1990's several authors studied the distribution of \ep\ for BATSE GRBs
(e.g. \cite{mall95,brai01}).
They reached the conclusion that \ep\ peaks around 200 keV
with few GRBs having \ep\ below 50 keV.
Recently Heise et al. (2001)  discovered short transients in the
Wide Field Cameras of BeppoSAX, which had little or no emission in the GRBM,
at energies above 40 keV. These events were called X-Ray Flashes (XRFs).
Kippen et al. (2001) found 9 of these events in the untriggered BATSE data and
performed a joint fit of the WFC+BATSE data in order to derive \ep\ for
these XRFs. They find values ranging from 4 to 90 keV, much lower than
the average for BATSE triggered GRBs.

Based on the criterion \ep\ $\leq$ 50 keV, we find that our sample contains 9 soft bursts
from a total of 33 long GRBs. While their \ep\ are not well constrained, we consider
that these 9 events certainly have \ep\ lower than 50 keV.
This percentage is comparable with 12 GRBs out of 22 with \ep $\le 50$ keV
in the GINGA sample (\cite{stro98}) and with 17 out of 66 GRBs in the
BeppoSAX/WFC sample (\cite{heis01}).
Thus, following the lead of GINGA and BeppoSAX, HETE-2 confirms
the existence of soft GRBs (with \ep\ lower than 50 keV).
The connection of these soft GRBs with the population of ``classical''
GRBs with \ep\ of a few hundred keV is discussed in the next section.

\subsection{X-ray rich GRBs and the Hardness-Intensity Correlation} \label{hic}

\begin{figure}
\resizebox{\hsize}{!}{\includegraphics{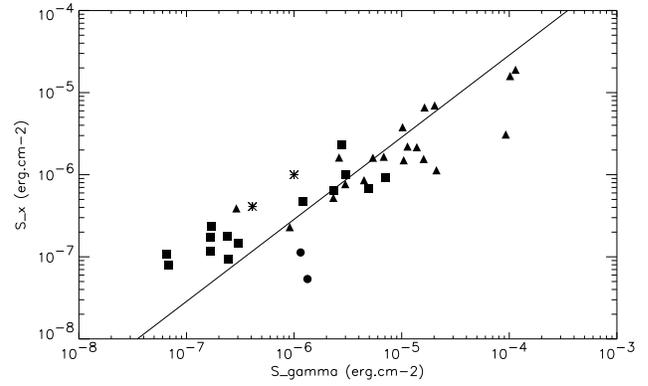}}
\caption{Fluence in the energy range 7-30 keV as a function of the
fluence in the range 30-400 keV.
Dark triangles show group A GRBs (see text) and empty triangles
group B bursts.
The solid line indicates events of constant hardness,
the spectral hardness is higher below the line.
The two stars indicate the position of the two most X-ray rich GRBs detected with the GRBM
on BeppoSAX (see text).}
\label{fluence}
\end{figure}

Since FREGATE provides, for the first time,
continuous coverage from 7 to 400 keV with
a single instrument, it is ideally suited to study the question of
whether events observed at low energies have the same properties
as the classical GRBs observed at higher energies.
In a first attempt to understand the possible connection between
soft GRBs and classical GRBs, we use the fluence/fluence diagram
plotted in Fig.~\ref{fluence}.
It is clear from this figure that there is no gap between the classical
GRBs and the soft GRBs.
Despite the small number of events, Fig.~\ref{fluence} suggests
a continuous evolution of GRB hardness with intensity.
This is the well known hardness-intensity correlation (hereafter HIC),
but FREGATE shows that this correlation extends over 3 orders of
magnitude in fluence.

In addition to FREGATE GRBs, Fig.~\ref{fluence} also displays two stars indicating
the position of the two most X-ray rich GRBs detected by the GRBM on BeppoSAX: 
GRB981226 on the left (\cite{fron00b}) and GRB990704 (\cite{fero01}). 
\cite{fron00b} say that ``GRB981226 has the weakest
gamma-ray peak flux detected with the BeppoSAX GRBM''. 
Fig.~\ref{fluence} shows that there is room for faint X-ray rich events which
are too faint for the GRBM and could only be seen by the WFC.
This confirms the results of \cite{kipp01} and clarifies the link between the X-ray rich GRBs 
detected by the GRBM+WFC on BeppoSAX and the XRFs detected only by the WFC.

\begin{figure}
\resizebox{\hsize}{!}{\includegraphics{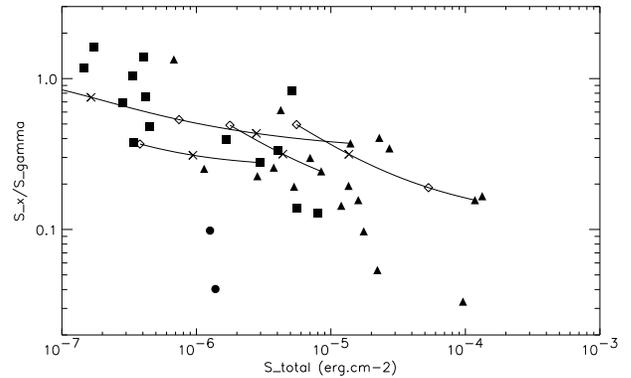}}
\caption{X-ray richness vs the total fluence. This plot shows the
hardness-intensity correlation observed by FREGATE. 
The four lines indicate how GRB010921, GRB020124, GRB020813, and GRB021004 
would evolve on this diagram if their redshifts were increased from the measured value
(z = 0.45, 3.2, 1.25, and 2.31 respectively) to z=10. 
For increasing redshifts, these GRBs move towards the upper left corner of the figure
(their fluence decreases and their X-ray richness increases).
Redshifts 1 and 5 are marked with crosses and redshifts 2 and 10 with empty squares.}
\label{durete}
\end{figure}

Another way to display the hardness-intensity correlation is
given in Fig.~\ref{durete} which shows the inverse of the hardness (defined as
the ratio of the fluence in the range 30-400 keV to the fluence
in the range 7-30 keV) as a function of the total fluence (in the range 7-400 keV).

In order to give a more quantitative statement on the significance of
this correlation we computed the average hardness ratio for the brightest 16 GRBs
and the faintest 16 GRBs in our sample (excluding the two short/hard GRBs). 
Taking into account only the
statistical errors (which are dominant) we find log(\lxlg) = -0.79 $\pm$ 0.07
for the bright GRBs and log(\lxlg) = -0.25 $\pm$ 0.07 for the faint GRBs.
This difference is significant at the 5.5 sigma level,
and it confirms the findings of other instruments
(\cite{atte94, mall95, deza97, atte00}).
A linear fit of the correlation between the soft fluence (7-30 keV) and
the hard fluence (30-400 keV) gives the following relation:

$$S_x = 3.2^{+2.7}_{-1.5} ~10^{-3} \times S_{\gamma} ^{0.643 \pm 0.046}$$

This fit should not however be taken too literally because the apparent deficit
of faint hard GRBs could be a selection effect caused by the small number of photons 
of these bursts. The lack of bright X-ray rich GRBs is real.

In the past the origin of the hardness-intensity correlation in GRBs
has been attributed to cosmological effects or to an intrinsic
hardness-luminosity correlation.
We now discuss these two interpretations using the FREGATE data.

Fig.~\ref{durete} plots the evolution of four GRBs with known redshift
(GRB010921 \cite{rick02}; GRB020124 \cite{hjor03}; GRB020813 \cite{pric02,fior02};
and GRB021004 \cite{chor02,sava02,cast02b}).
with the redshift (up to z=10) on a fluence/hardness diagram.
The spectra of these 4 GRBs were fit
with a Band function having the same alpha and \eo\ as the cutoff power law fit
and  $\beta = -2.3$.
For this study we used a Band function because it was not appropriate
to neglect the high energy
spectral index which plays an important role for GRBs at high redshift.
We chose $\beta = -2.3$ because it is the average value found by \cite{pree00}.
For these computations we asssumed a flat universe with $\Omega_0$ = 0.7, $\Omega_\Lambda$ = 0.3
and H$_0$ = 65 km sec$^{-1}$ Mpc$^{-1}$.

\begin{figure}
\resizebox{\hsize}{!}{\includegraphics{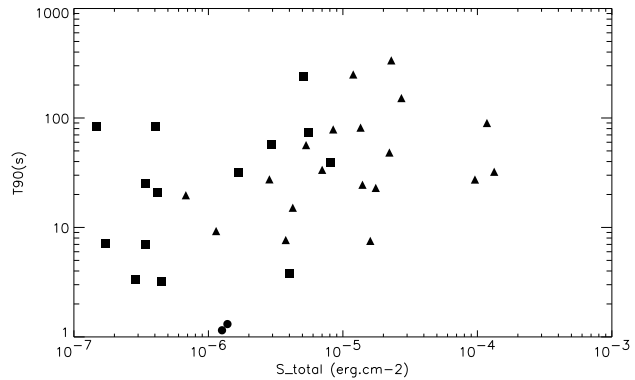}}
\caption{GRB duration (T$_{90}$) as a function of the total fluence.
This figure shows that faint GRBs are not significantly longer than
bright events.}
\label{duree}
\end{figure}

Figure~\ref{durete} shows that cosmological effects could in principle explain the
observed correlation.
In this case, however, we would also expect a significant time dilation
of the soft GRBs.
Fig.~\ref{duree} plots the duration T$_{90}$ as a function of the total fluence.
It shows that there is no significant time dilation of the faint GRBs.
We note here that because the
widths of the peaks in the time histories of GRBs -- and the durations
of GRBs -- are shorter at higher energies (e.g., \cite{feni95}), this
partly (but only partly) compensates for the time dilation due
to the cosmological redshift.  
GRB durations go approximately like E$^{-0.4}$, so
that this effect {\it shortens} the observed durations of GRBs at a redshift
z = 10 relative to the durations of GRBs at a redshift z =1 by a factor
of about [(1+10)/(1+1)]$^{0.4}$ = 2. 
Time dilation would be expected to {\it increase} the duration of the bursts by
[(1+10)/(1+1)] = 5.5.  Thus, overall, one expects bursts at
high redshifts to be longer by a factor of only about 2.7.  
Still, Fig.~\ref{duree} does not support such a dependence.

While it is always possible to invoke GRB evolution to produce intrinsically
shorter GRBs at high redhifts, we consider that our observations do not favor
the interpretation of the HIC purely in terms of cosmological effects.
Finally, we note
that Amati et al. (2002) find a correlation between the intrinsic (redshift corrected) 
\ep\ of 12 GRBs with known redshifts, and  E$_{52}$, their isotropic-equivalent energy radiated in gamma-rays,
in units of 10$^{52}$ ergs:
\ep\ = 100 E$_{52}^{0.52}$ keV.
This correlation, if it extends over a sufficient range of redshifts could
certainly explain the hardness-intensity correlation we observe.
With this interpretation, the HIC would be the reflection of a more
fundamental correlation between the radiated isotropic-equivalent energy and
the spectral hardness in GRBs. Our observations suggest that this correlation 
could include the X-ray rich GRBs. 
If X-ray rich GRBs are intrinsically fainter, we should also expect them to be closer
on average than bright GRBs. We discuss this issue in the next section.

\subsection{The distance and nature of X-ray rich GRBs} \label{xnature}


A quick survey of the literature shows that X-ray rich GRBs have no
or very faint optical afterglows. 
The only tentative identification of an optical afterglow is reported by \cite{fruc02c} 
for GRB020410: a faint source is seen with the STIS on the HST at V=25.4 on May 8 and 
V=26.9 on June 14. This identification is not certain 
however due to the possible confusion with a field supernova unrelated to the GRB.

X-ray rich GRBs, on the other hand, seem to have X-ray afterglows (e.g.  GRB981226,
GRB990704, GRB011030, GRB020410 or GRB020427) and for some of them radio afterglows (GRB981226, GRB011030).
Thanks to the good localization capabilities of Chandra, the host galaxies of GRB981226, GRB011030 and 
GRB020427 have been identified (\cite{frai99,fruc02a,cast02a,fruc02b}), but their redshifts have 
not yet been measured. 
To summarize, we still have no measure of the distance of an X-ray rich GRB.\footnote{It is probably useful to note here 
that GRB011211 which was initially classified as an X-ray rich GRB was then reclassified as a normal GRB
\cite{fron02}. The afterglow of this GRB and its redshift z=2.14 were for some time considered as strong
arguments in favor of similar distance scales for normal GRBs and X-ray rich GRBs.}

The nature of X-ray rich GRBs can be addressed from the theoretical or from an empirical
point of view. From the theoretical point of view, we note that the model of internal shocks predicts
that X-ray rich GRBs could be produced by fireballs with less efficient shocks (due to lower 
magnetic fields or to a lower contrast of the Lorentz factors within the ejecta) or
by {\it clean} fireballs (with a low baryon load) (\cite{zhan02,daig02,moch03}). 
In a clean fireball the {\it initial} Lorentz factor is higher but
the internal shocks take place at larger distances from the central source where the density
and the magnetic fields are smaller, leading to the emission of less energetic photons.
Further theoretical studies are needed to assess whether one of these conditions can also
explain the correlation found by \cite{amat02}, the lack of bright optical afterglows and the unusual
properties of the X-ray afterglows of X-ray rich GRBs (\cite{fron00b, fero01}).

From the empirical point of view, if we combine the 
evidence in this paper that (a) X-ray rich GRBs are not a separate
class of GRBs but represent an extension of the properties of "typical"
GRBs, and (b) the HIC correlation extends over three orders of
magnitude in fluence and applies to X-ray rich GRBs with 
the conclusion of \cite{frai01} that the total energy of GRBs is roughly constant,
we are led to the conclusion that the jet opening angle of X-ray rich GRBs are substantially
larger than the jet opening angle of "typical" GRBs.
Additional observations are clearly required to understand
the role of the progenitor and/or its environment in shaping
the properties of the prompt GRB emission, particularly the peak energy. 

\section{Conclusions} \label{conclusion}

Our observations have two interesting consequences: they confirm that
the \ep\ distribution is broader than previously thought (\cite{mall95, pree00, brai00})
and they show that we do not see yet the faint end of the GRB distribution.
If we assume that the correlation found by Amati et al. (2002) extends
down to \ep\ as low as 20 keV, it would
imply that the isotropic-equivalent energy radiated by a GRB with \ep\ =
20 keV is about 80 times smaller than the isotropic-equivalent energy
radiated by a "typical" GRB with \ep\ = 200 keV.
If the conclusion of
Frail et al. (2000) that the total energy of GRBs is roughly constant,
it implies that the jet opening angle of X-ray rich GRBs are
substantially larger than the jet opening angle of "typical" GRBs.

Future work with HETE-2 will bring several advances in this field and
should contribute to our understanding of the population
of soft/faint GRBs. The continuously growing GRB sample of FREGATE
should provide better statistical evidence for the effects discussed
in this paper and additional clues about the possible differences between
bright and faint GRBs and about the nature of X-ray rich GRBs. 
Joint spectral analysis
with the WXM will allow more precise determinations of $\alpha$ and
\eo\ for X-ray rich GRBs. 
Finally, measuring the redshifts of a greater
number of GRBs detected by HETE-2 will allow us to test the extent of
the correlation between the spectral hardness of GRBs and their radiated
energy in gamma-rays.
X-ray rich GRBs also present an interesting challenge for future GRB
missions and for observers on the ground.
Future GRB missions will have to detect events which are much softer and
fainter than the typical GRB population sampled by BATSE.  
Observers on the ground are faced with events which
have fainter afterglows than the classical gamma-ray bursts.

To conclude we note that the joint detection of GRB010213 (with \ep = 2.5 keV, \cite{kawa03,saka03})
by WXM and FREGATE and of GRB020903 (with \ep = 0.9 keV, \cite{kawa03}) by WXM only 
demonstrate the existence of events which are even softer than the bulk of the X-ray rich GRBs discussed
in this paper.

\begin{acknowledgements}
The HETE-2 mission is supported in the US by NASA contract NASW-4690;
in France by CNES contract 793-01-8479; and in Japan in part by
the Minisitry of Education, Culture, Sports, Science and
Technology Grant-in-Aid 13440063.
KH is grateful for support under MIT contract SC-R-293291.
GP aknowledges support by the Italian Space Agency (ASI)
The authors acknowledge the support of the HETE-2 operation team.
The authors acknowledge the use of J. Greiner GRB page at http://www.mpe.mpg.de/~jcg/grbgen.html.
\end{acknowledgements}

\newpage

\newpage

\section{Appendix: Spectra of the 35 GRBs}
   \begin{figure}[hb]
   \includegraphics[angle=-90,width=8.0cm]{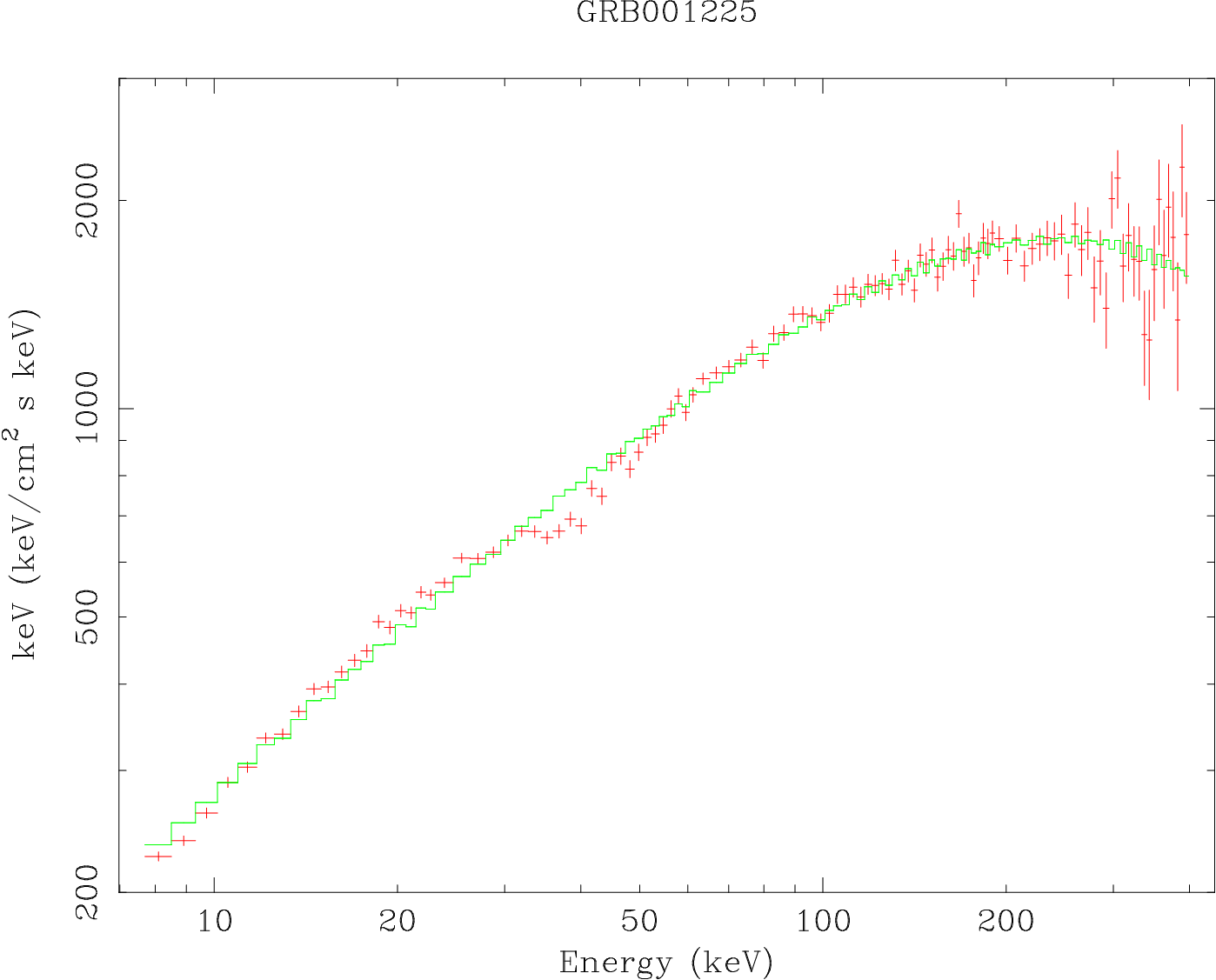} \vspace{0.5cm}

   \includegraphics[angle=-90,width=8.0cm]{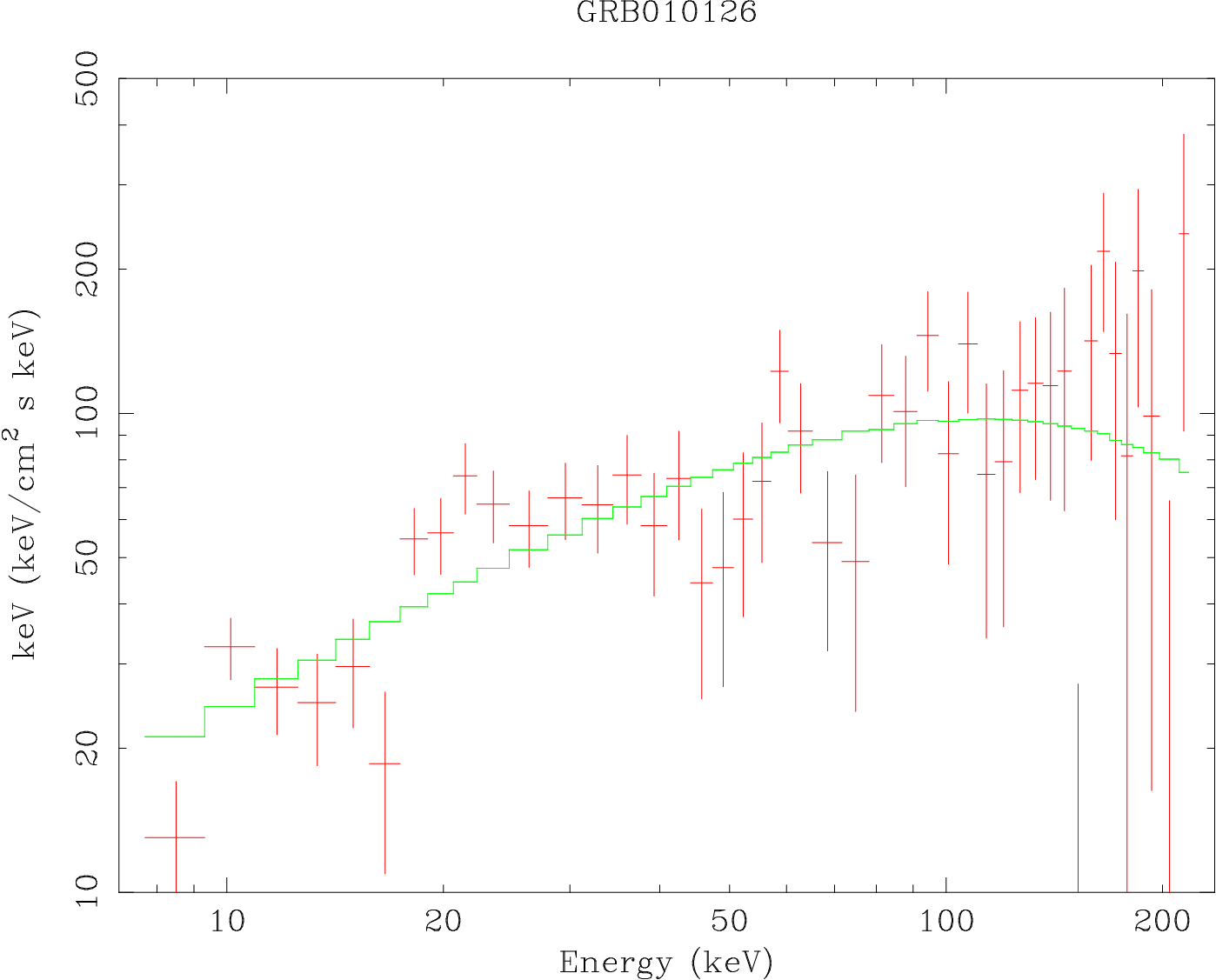} \vspace{0.5cm}

   \includegraphics[angle=-90,width=8.0cm]{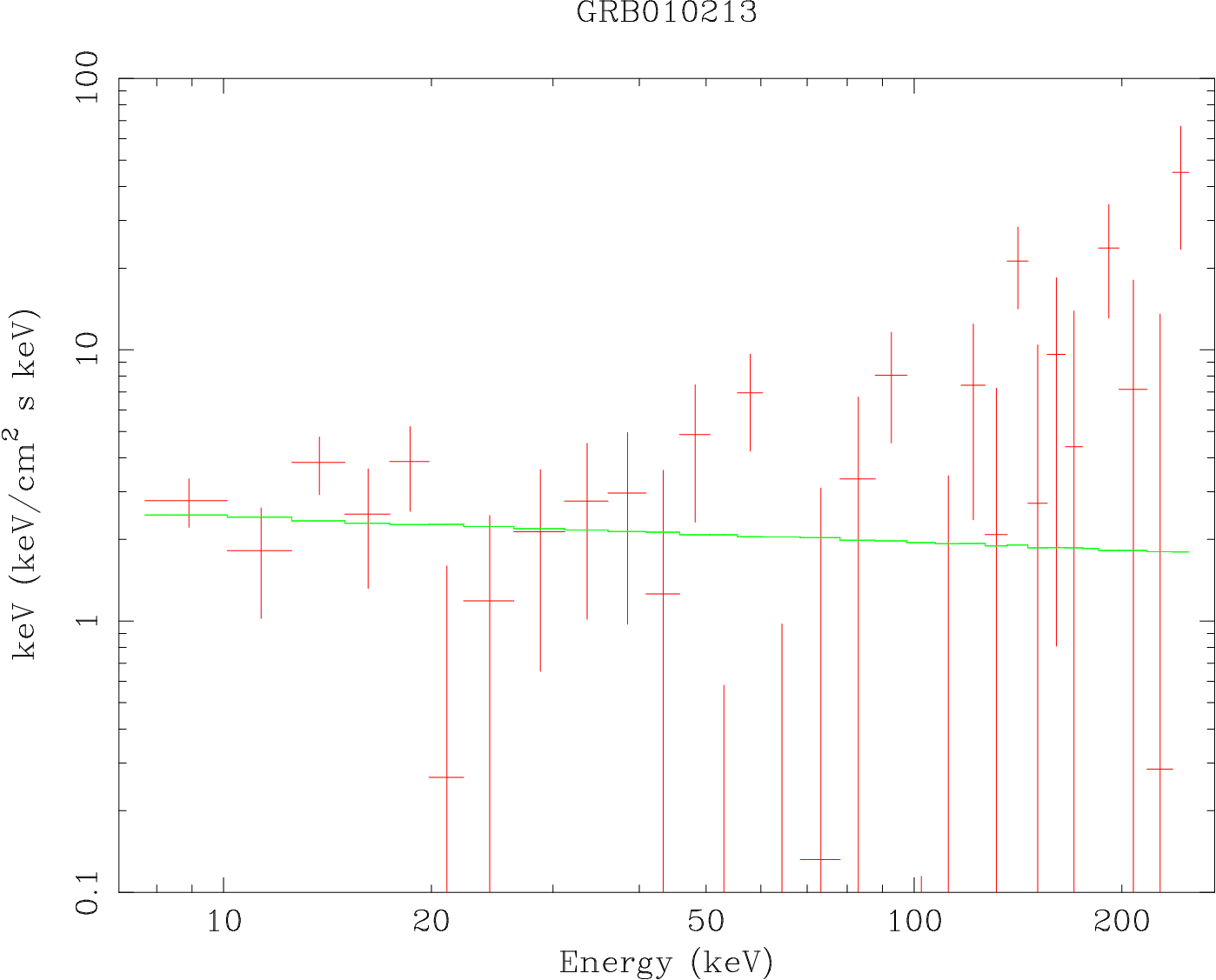} 
   \end{figure}

   \begin{figure}
   \includegraphics[angle=-90,width=8.0cm]{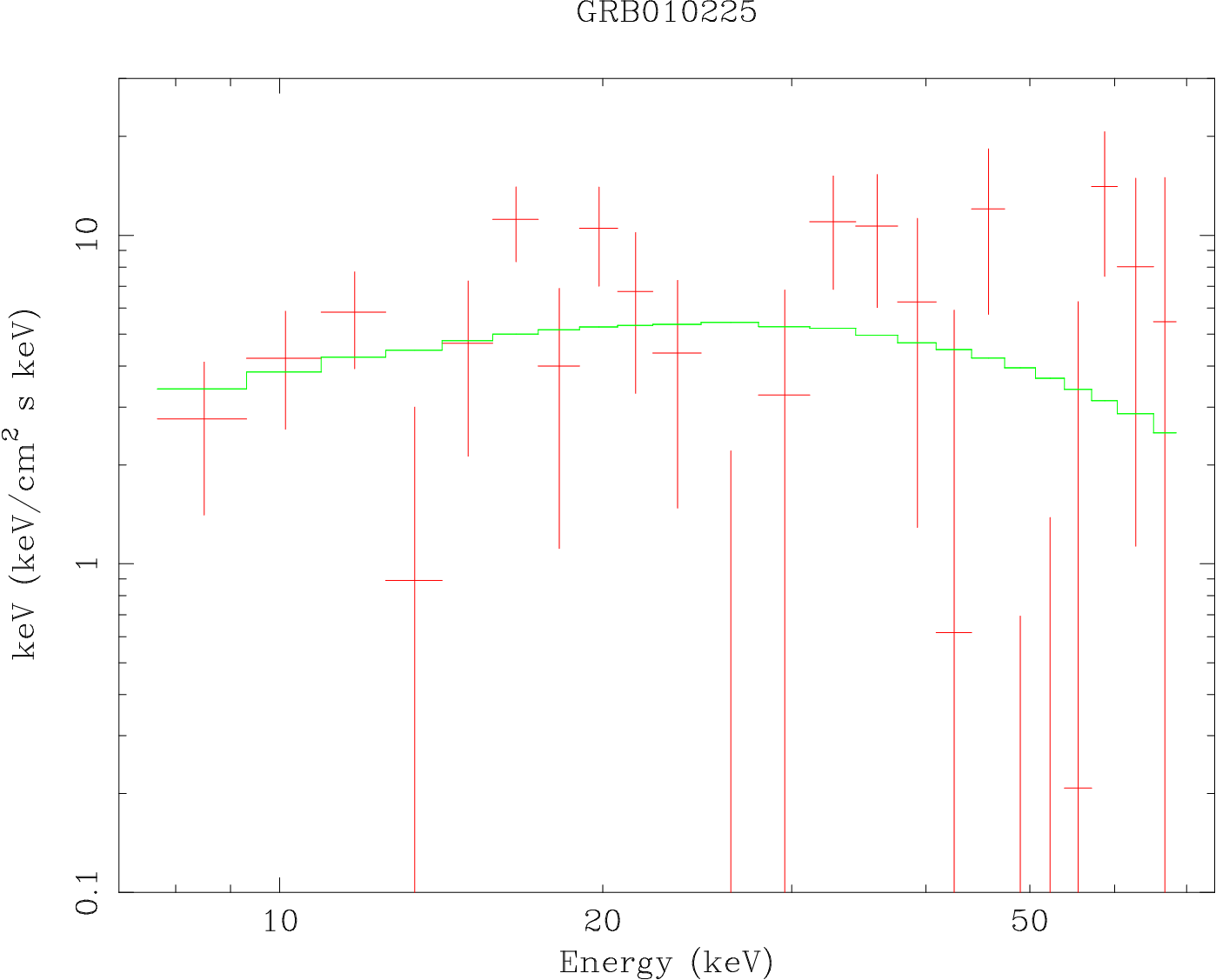} \vspace{0.5cm}

   \includegraphics[angle=-90,width=8.0cm]{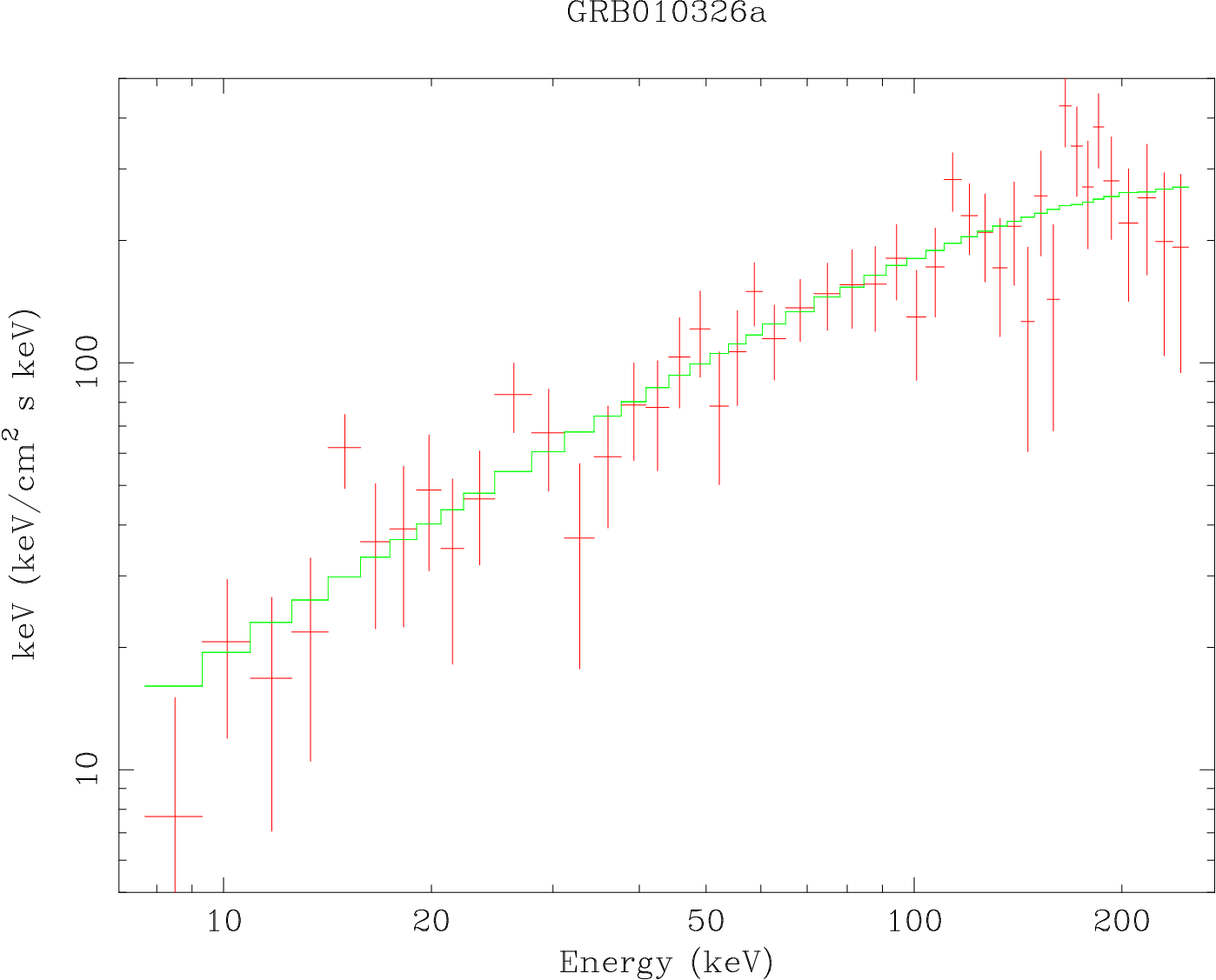} \vspace{0.5cm}

   \includegraphics[angle=-90,width=8.0cm]{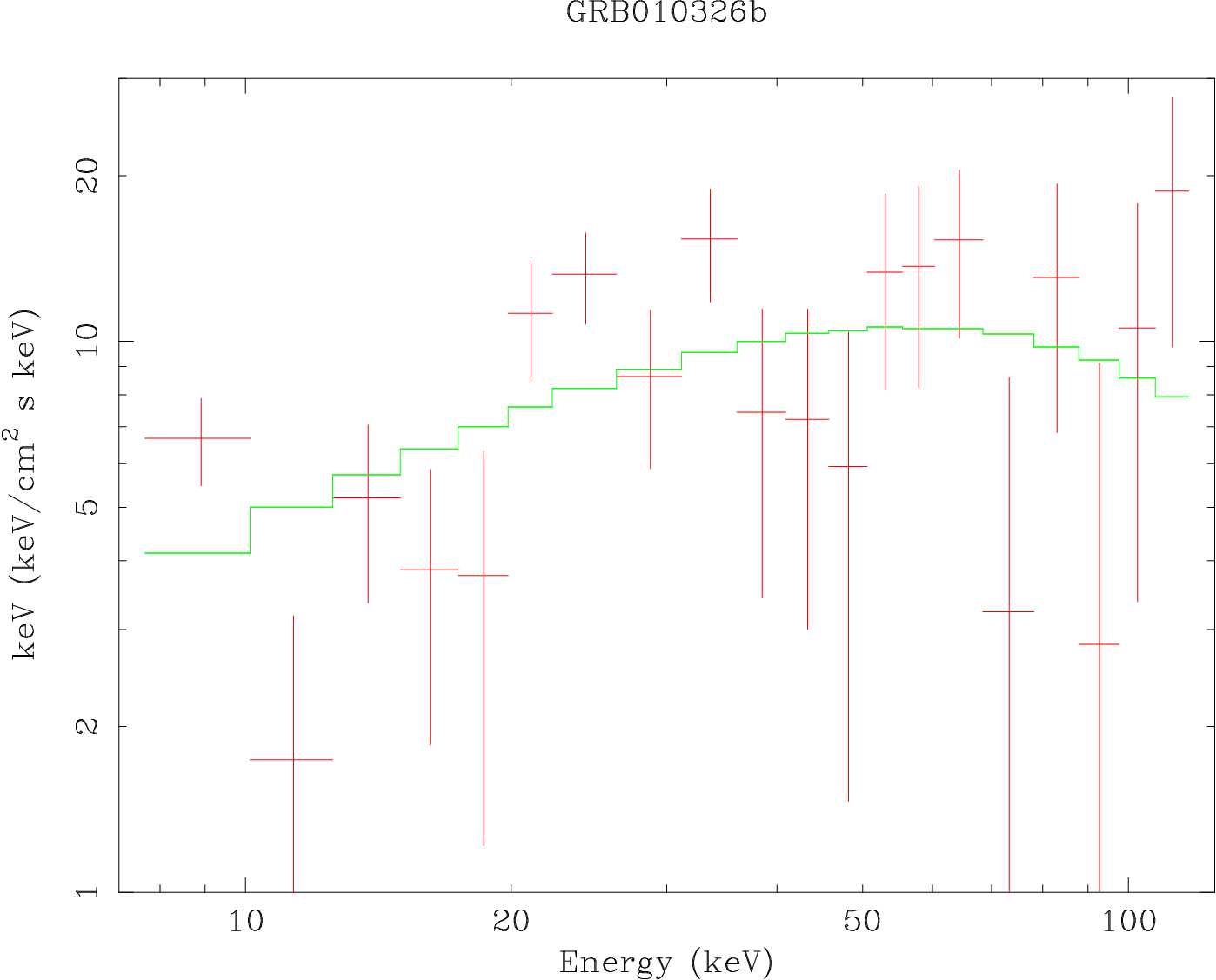} \vspace{0.5cm}

   \includegraphics[angle=-90,width=8.0cm]{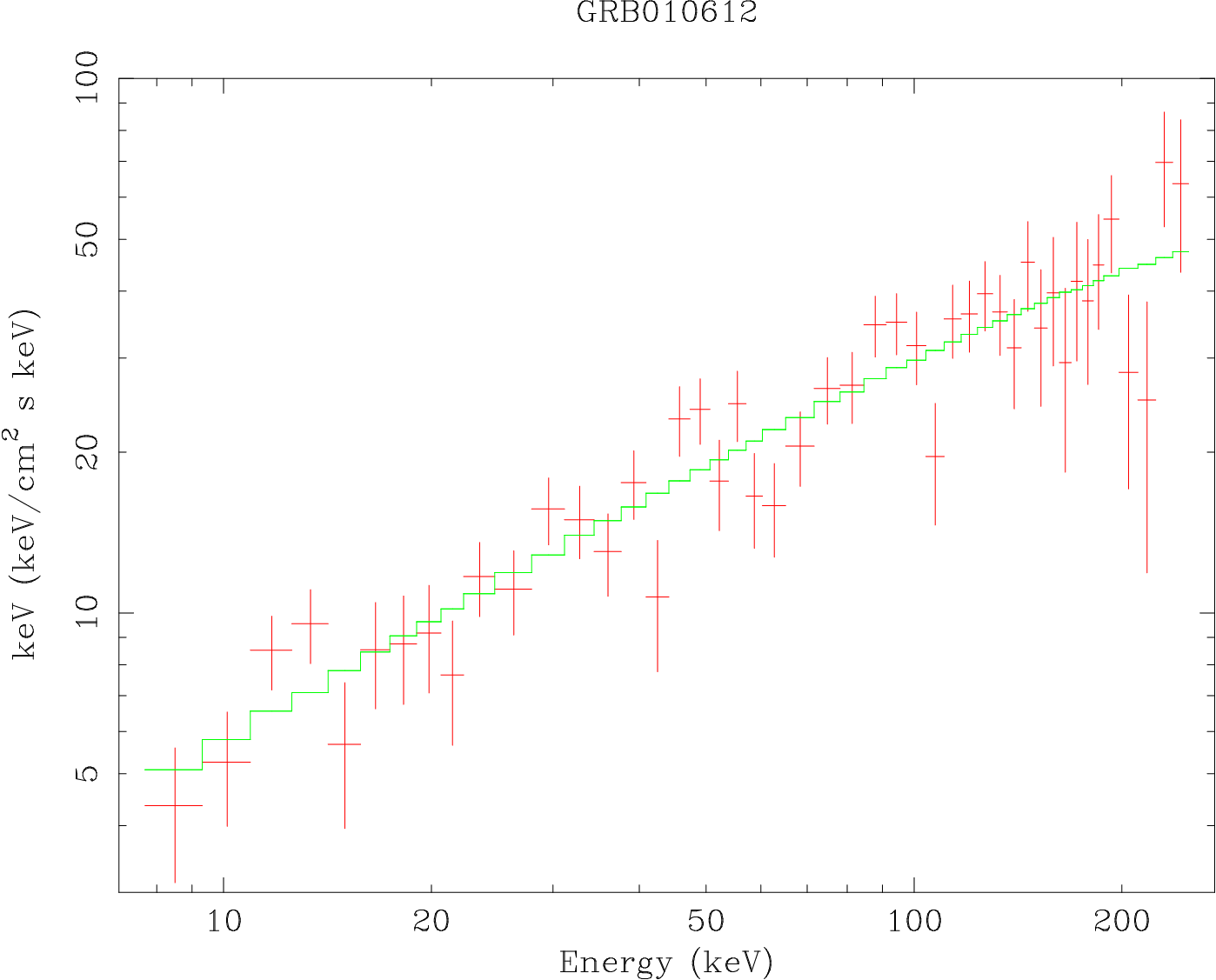} 
   \end{figure}

   \begin{figure}
   \centering
   \includegraphics[angle=-90,width=8.0cm]{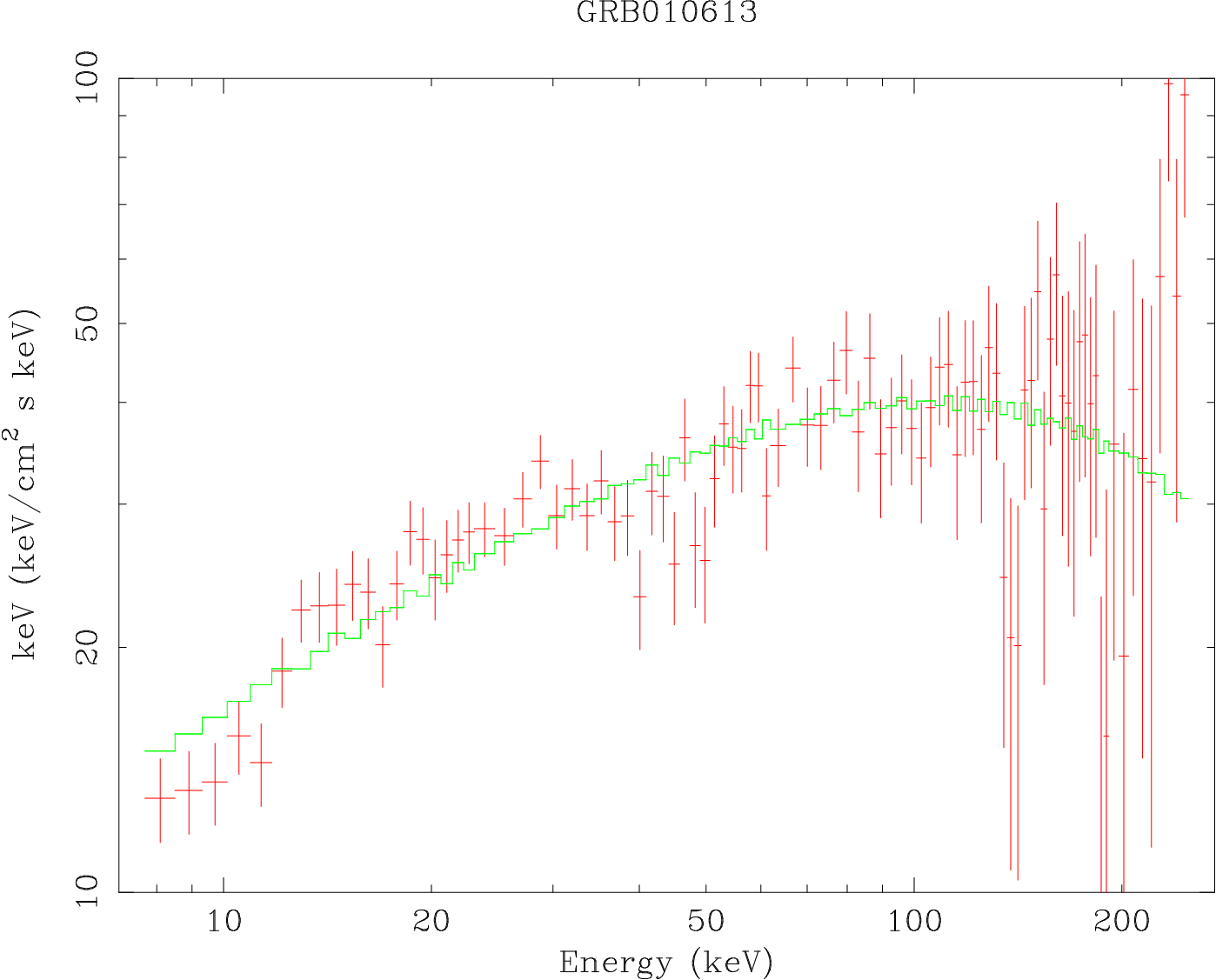} \vspace{0.5cm}

   \includegraphics[angle=-90,width=8.0cm]{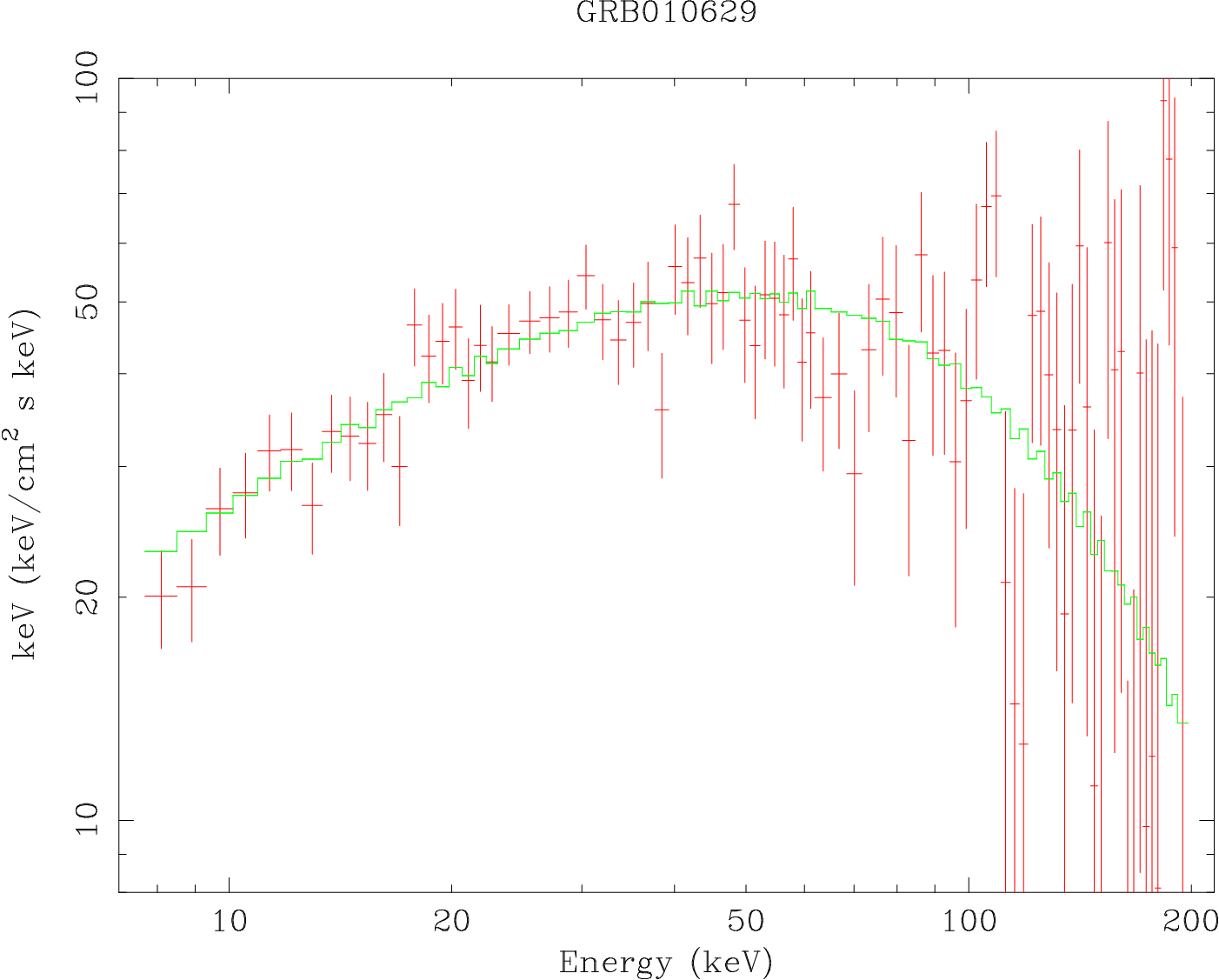} \vspace{0.5cm}

   \includegraphics[angle=-90,width=8.0cm]{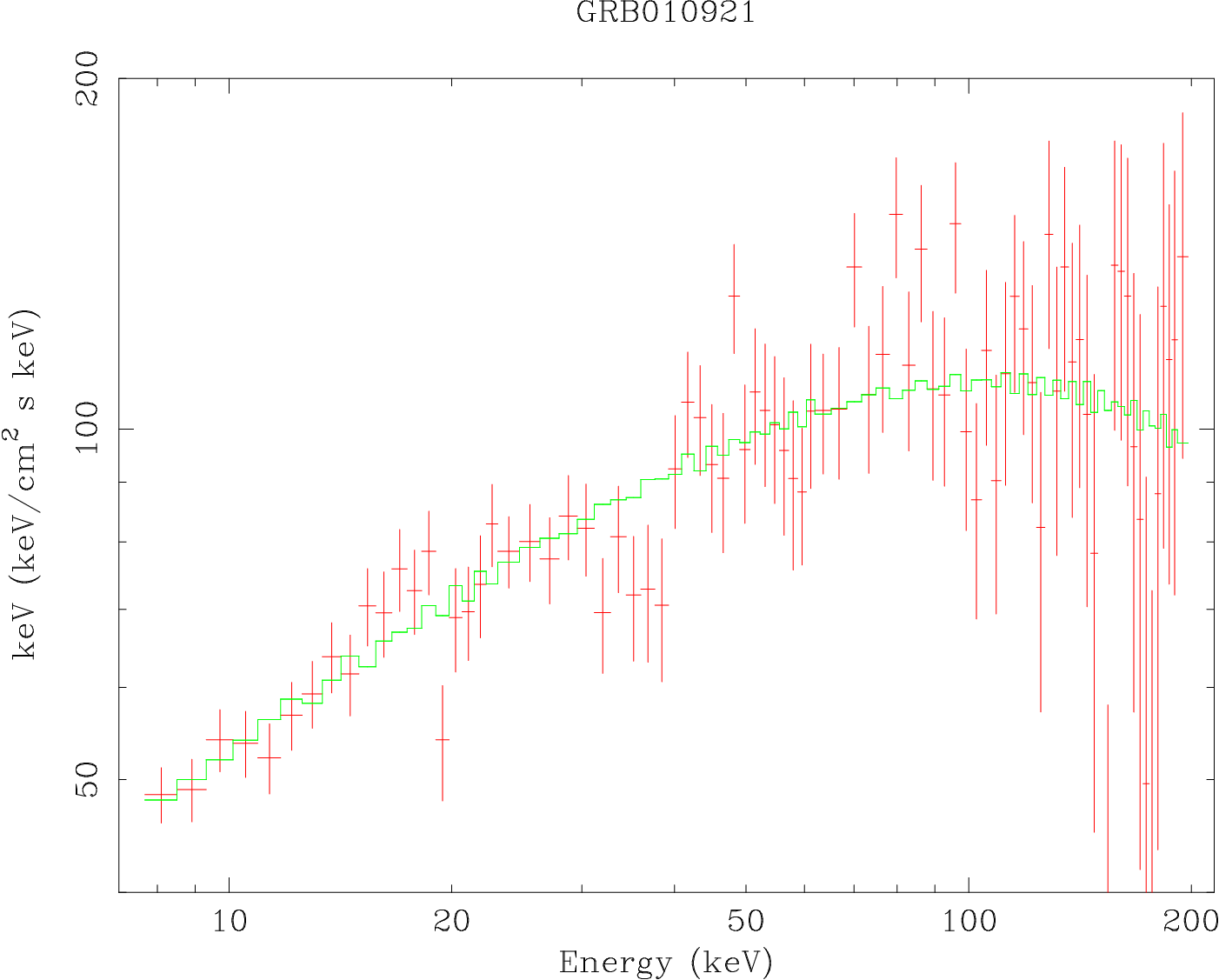} \vspace{0.5cm}

   \includegraphics[angle=-90,width=8.0cm]{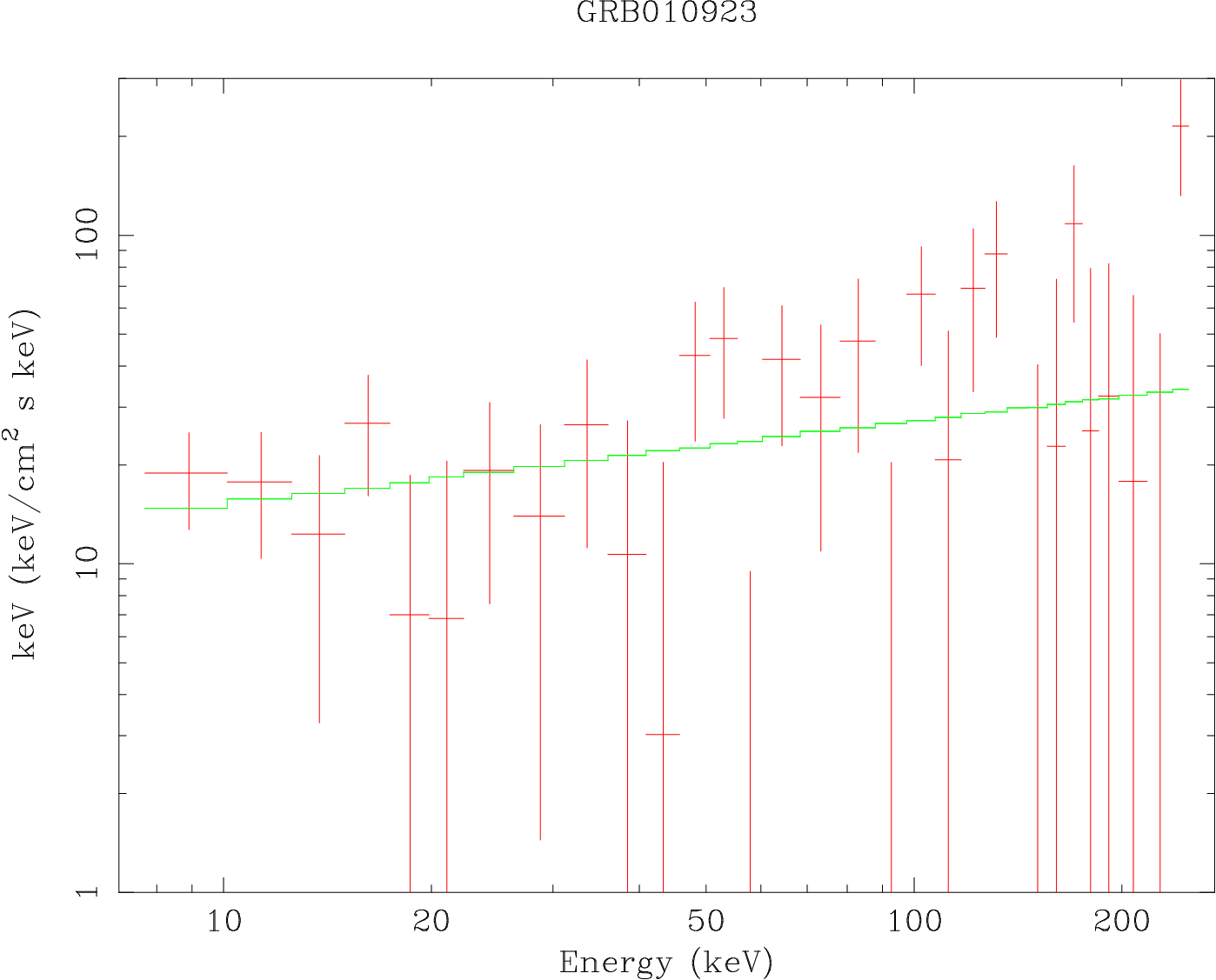}
   \end{figure}

   \begin{figure}
   \centering
   \includegraphics[angle=-90,width=8.0cm]{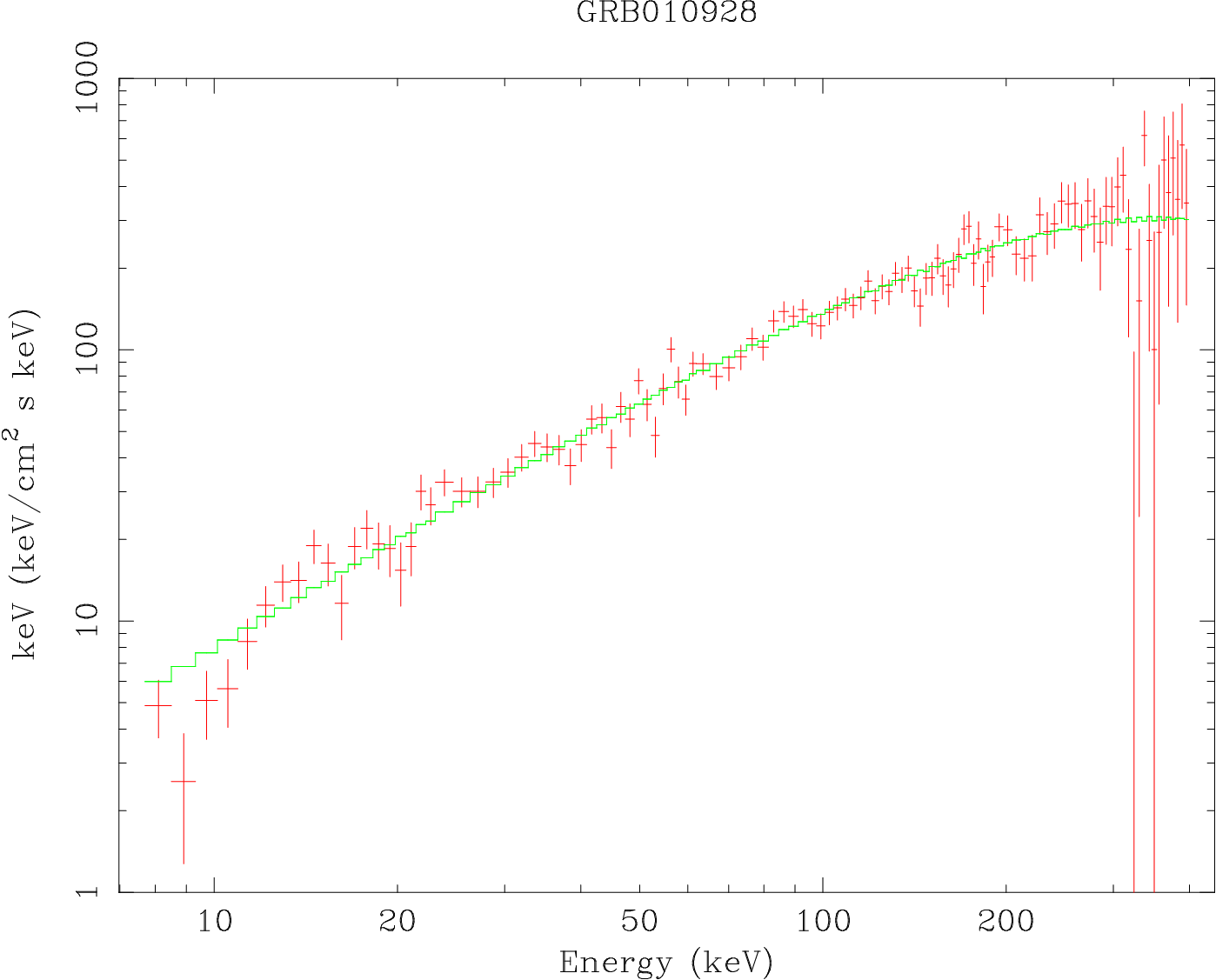} \vspace{0.5cm}

   \includegraphics[angle=-90,width=8.0cm]{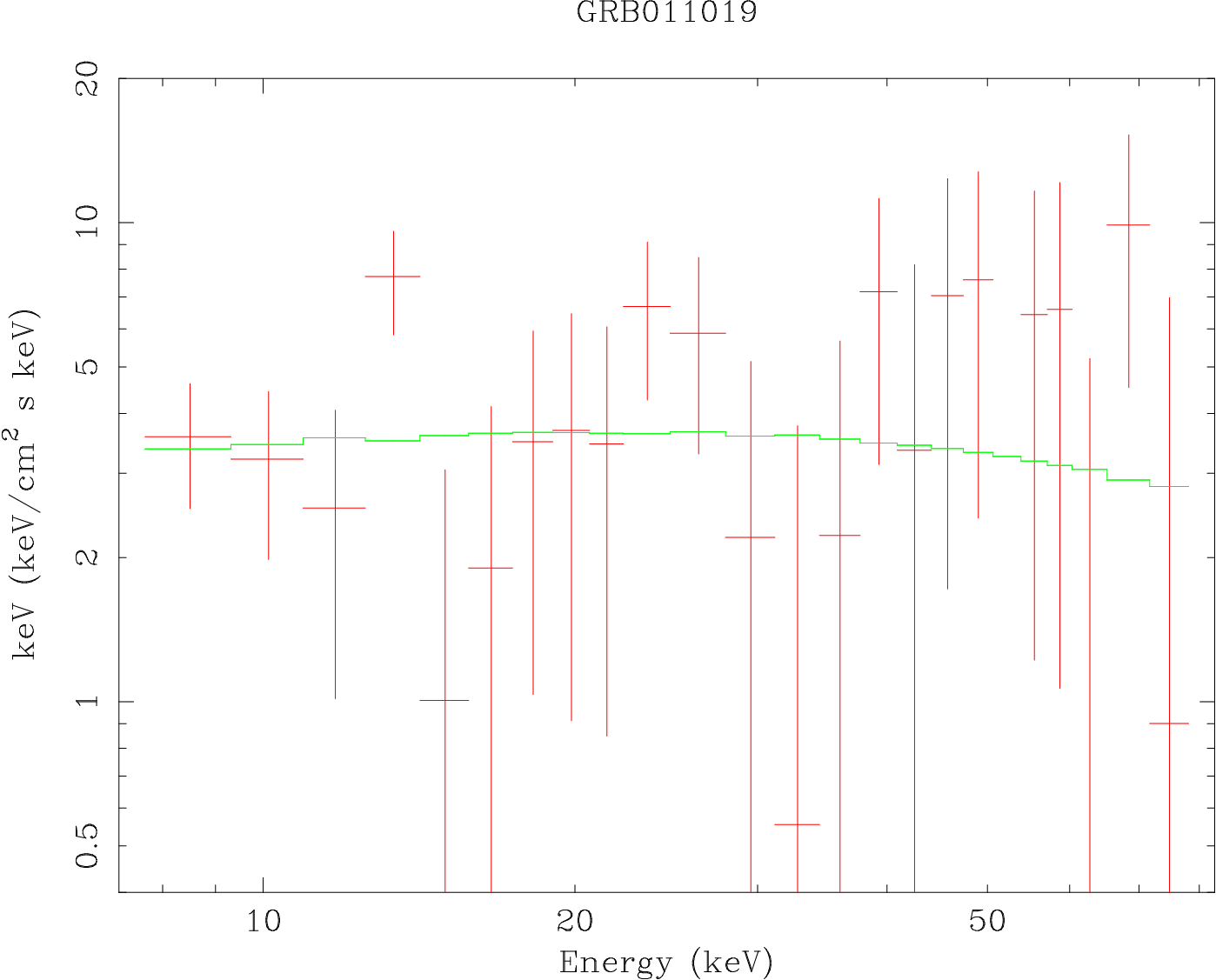} \vspace{0.5cm}

   \includegraphics[angle=-90,width=8.0cm]{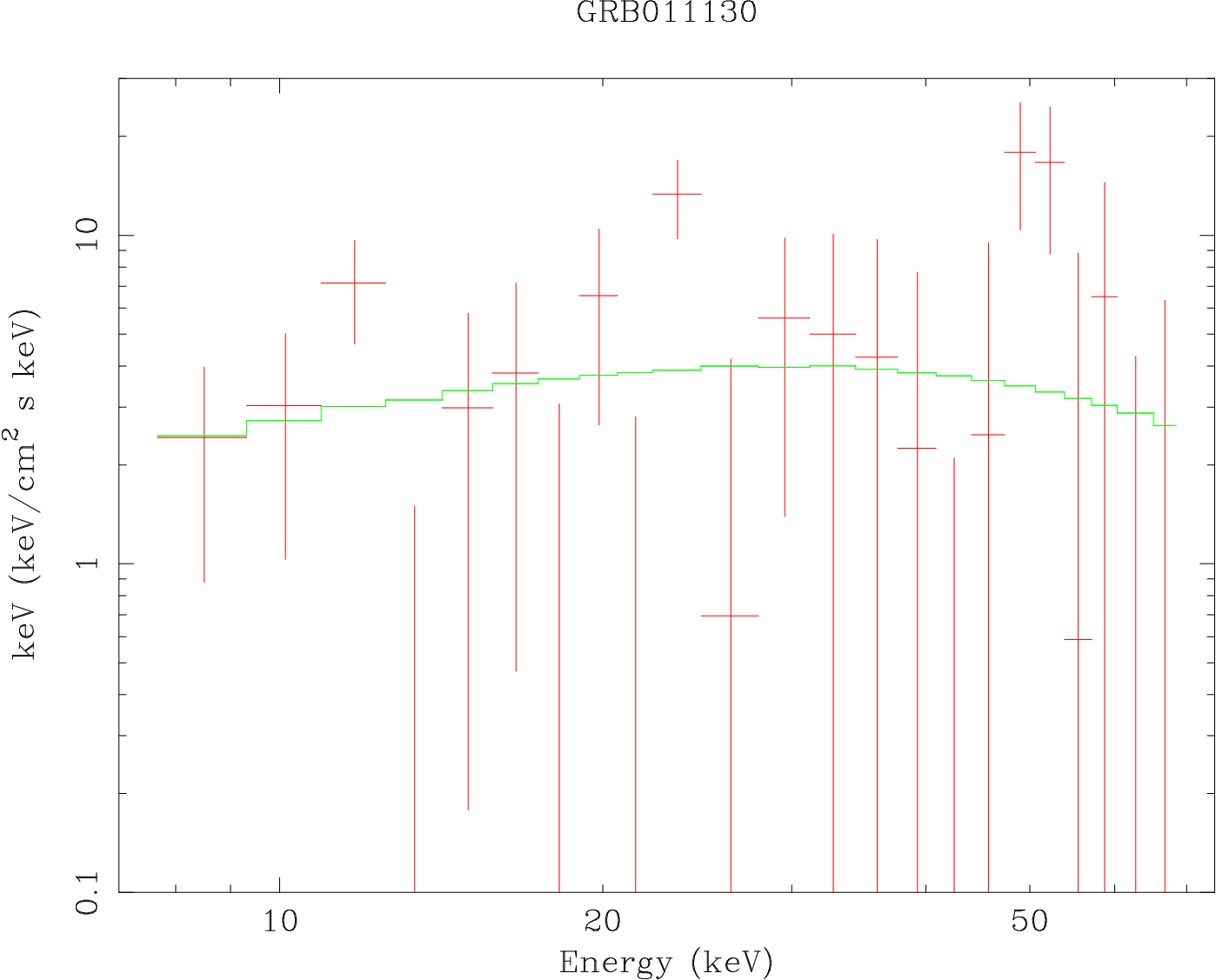} \vspace{0.5cm}

   \includegraphics[angle=-90,width=8.0cm]{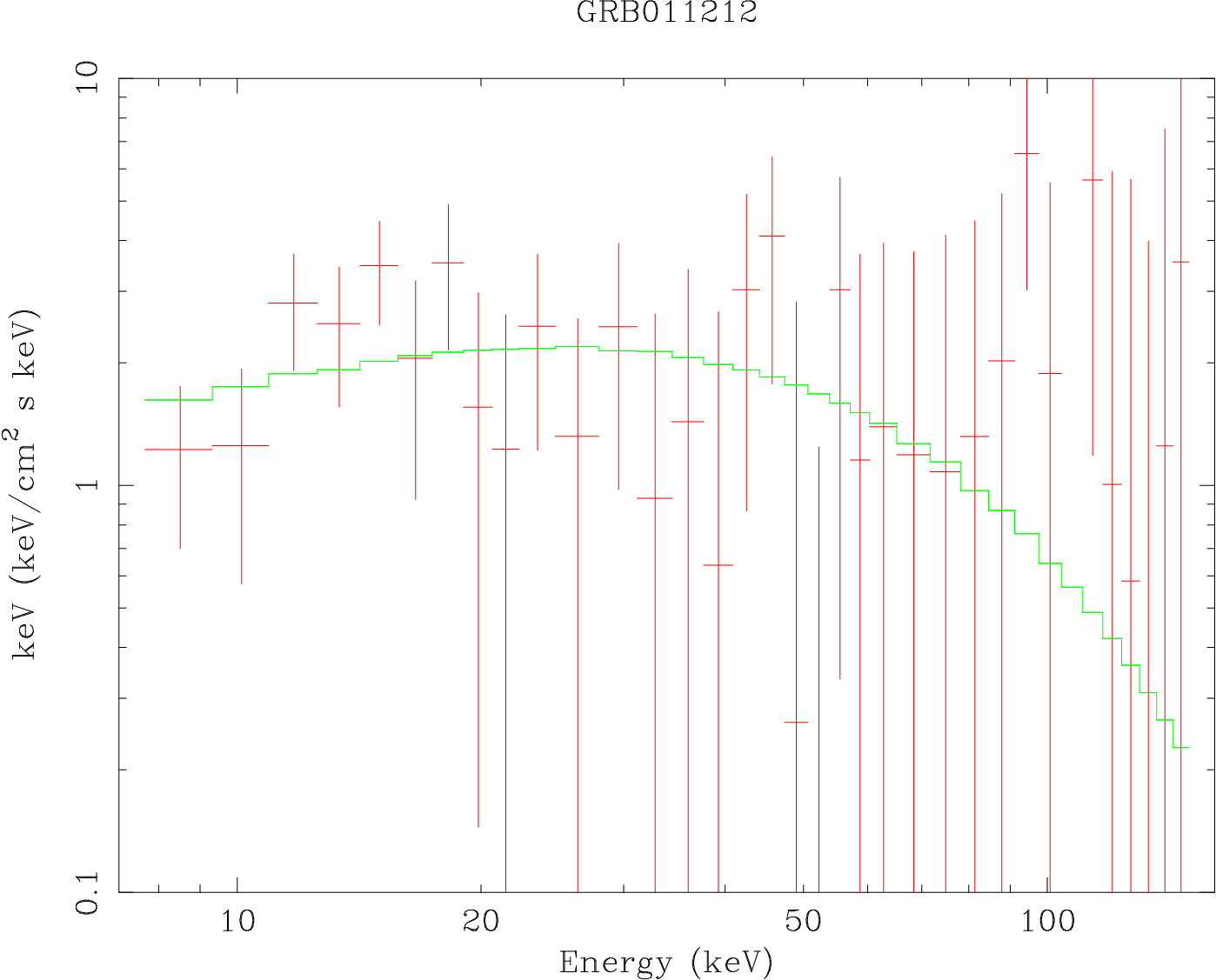} 
   \end{figure}

   \begin{figure}
   \centering
   \includegraphics[angle=-90,width=8.0cm]{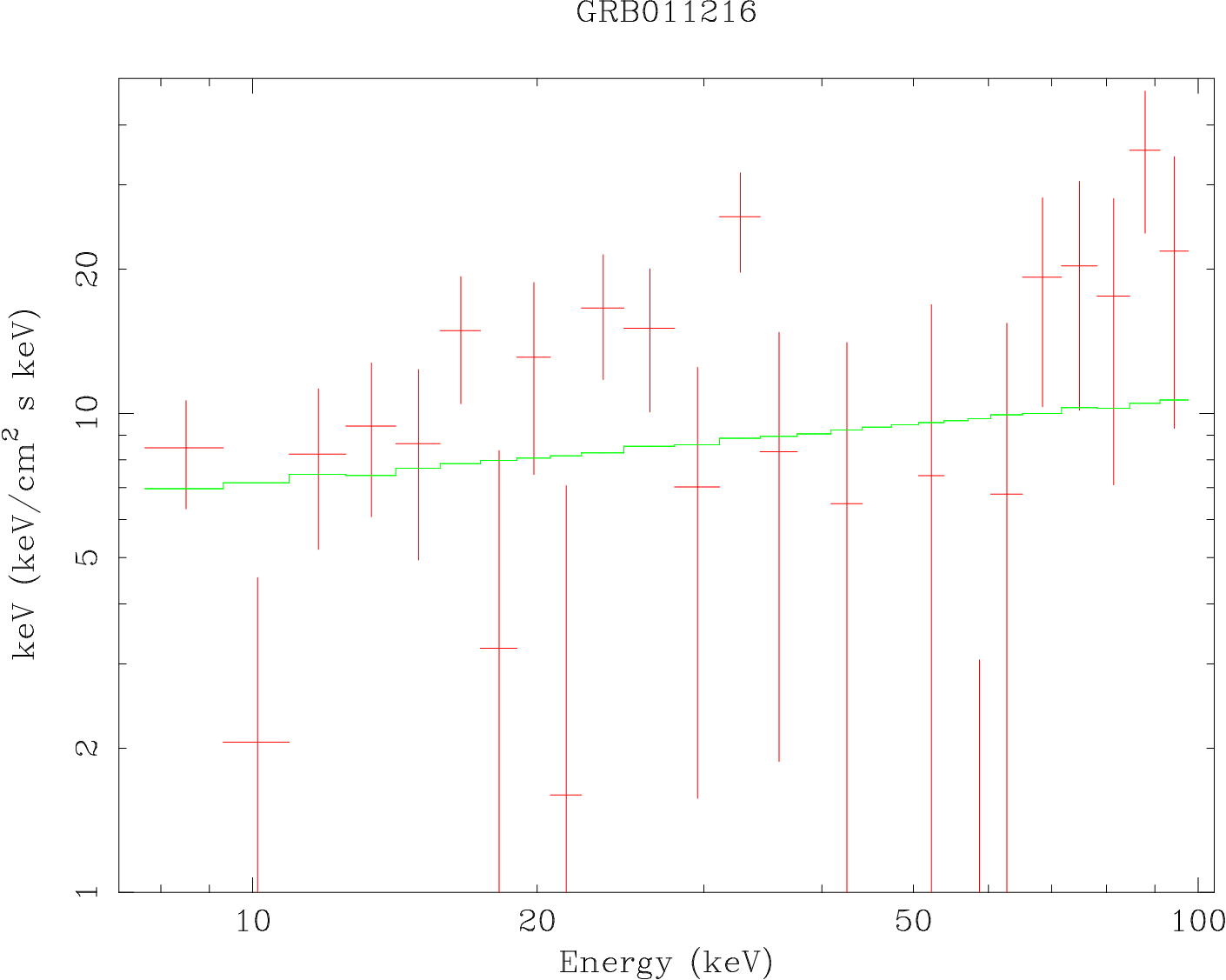} \vspace{0.5cm}

   \includegraphics[angle=-90,width=8.0cm]{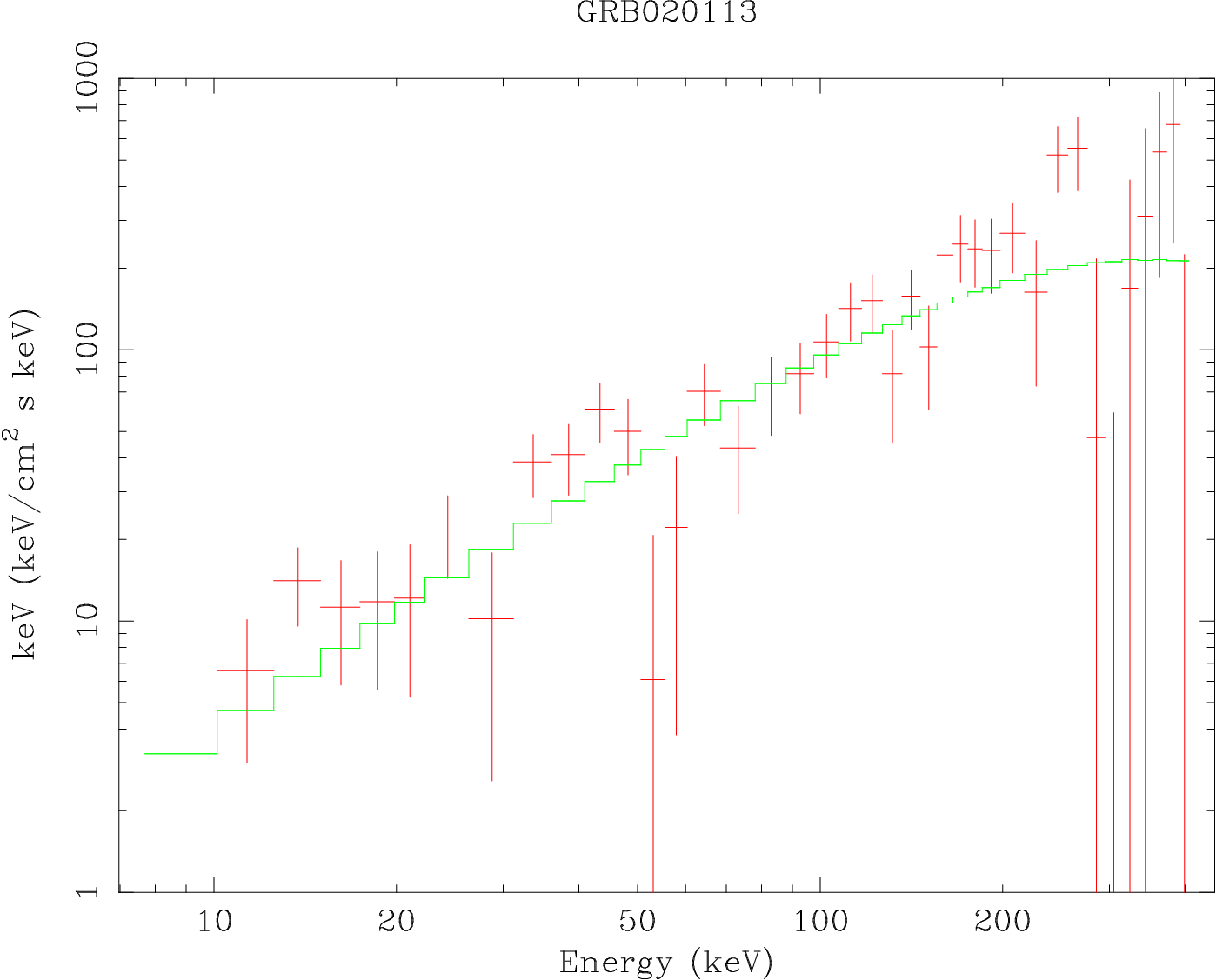} \vspace{0.5cm}

   \includegraphics[angle=-90,width=8.0cm]{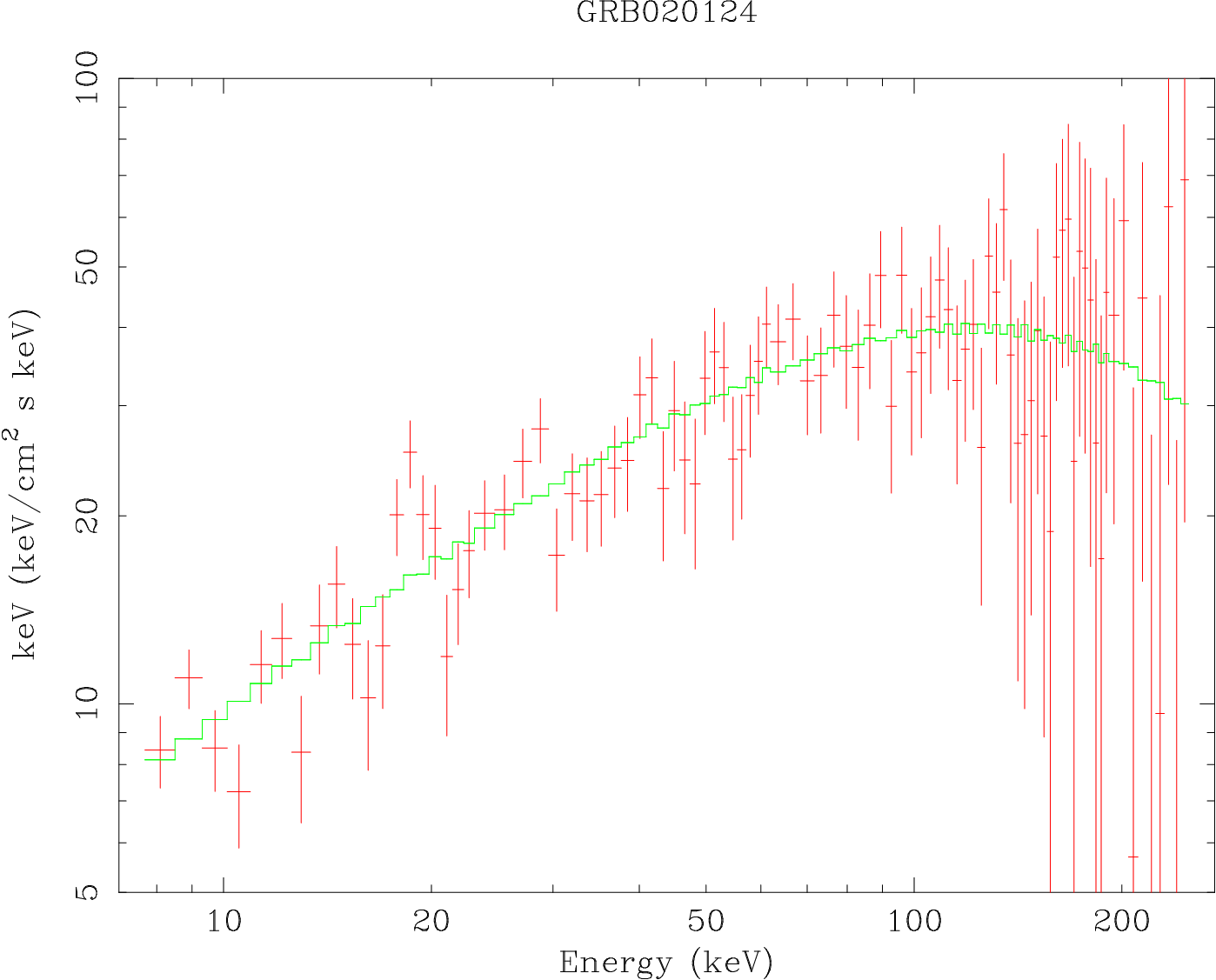} \vspace{0.5cm}

   \includegraphics[angle=-90,width=8.0cm]{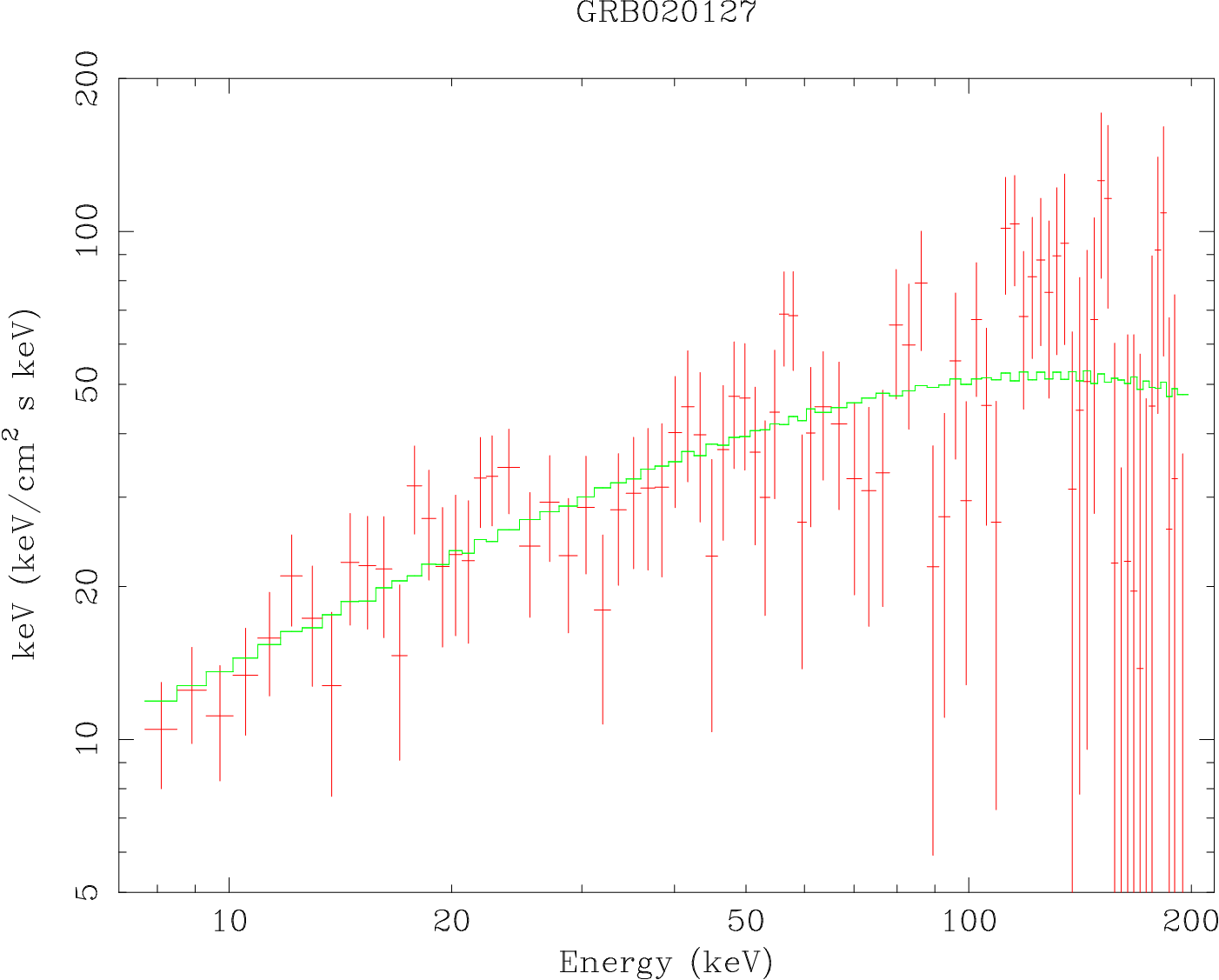}
   \end{figure}

   \begin{figure}
   \includegraphics[angle=-90,width=8.0cm]{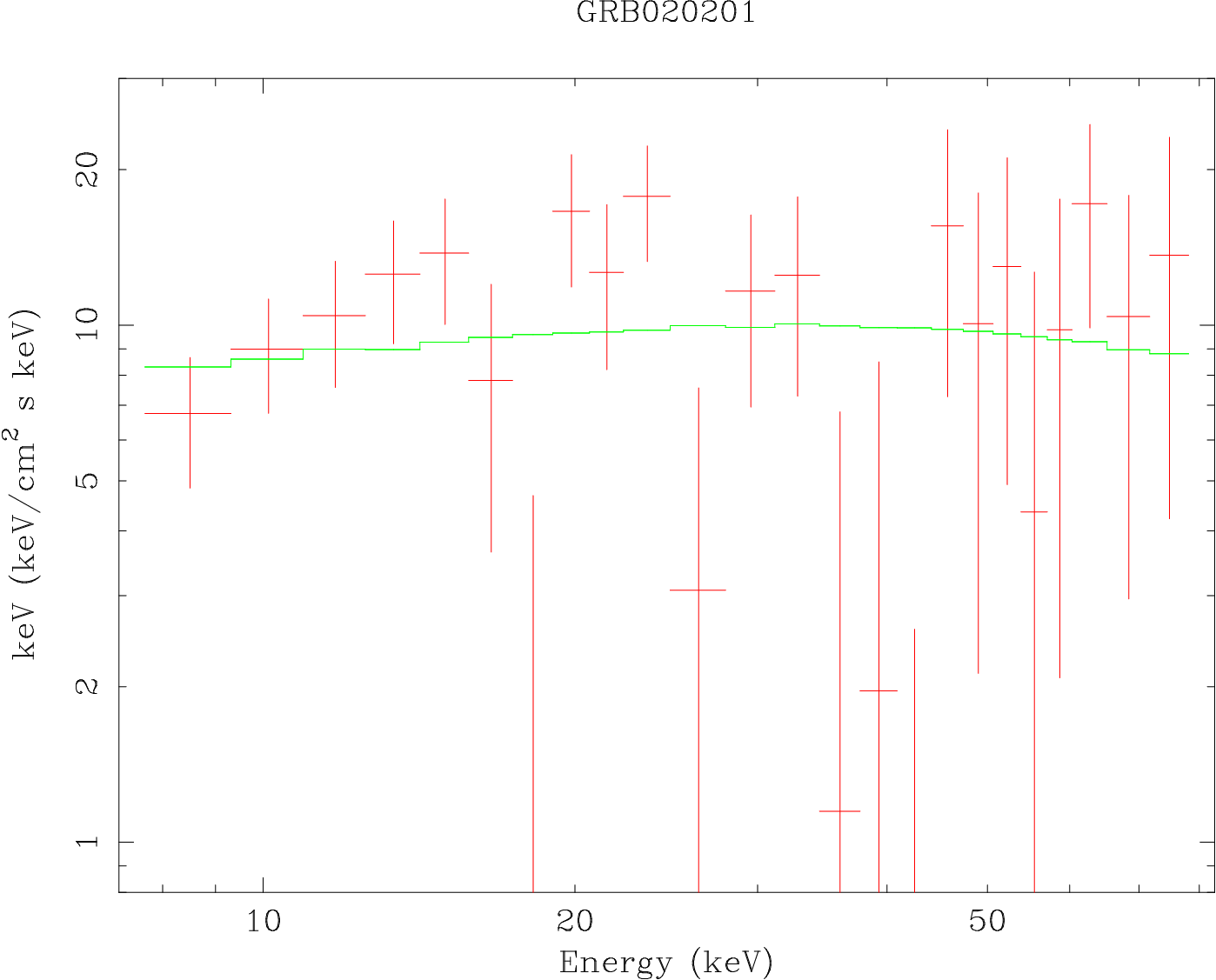} \vspace{0.5cm}

   \includegraphics[angle=-90,width=8.0cm]{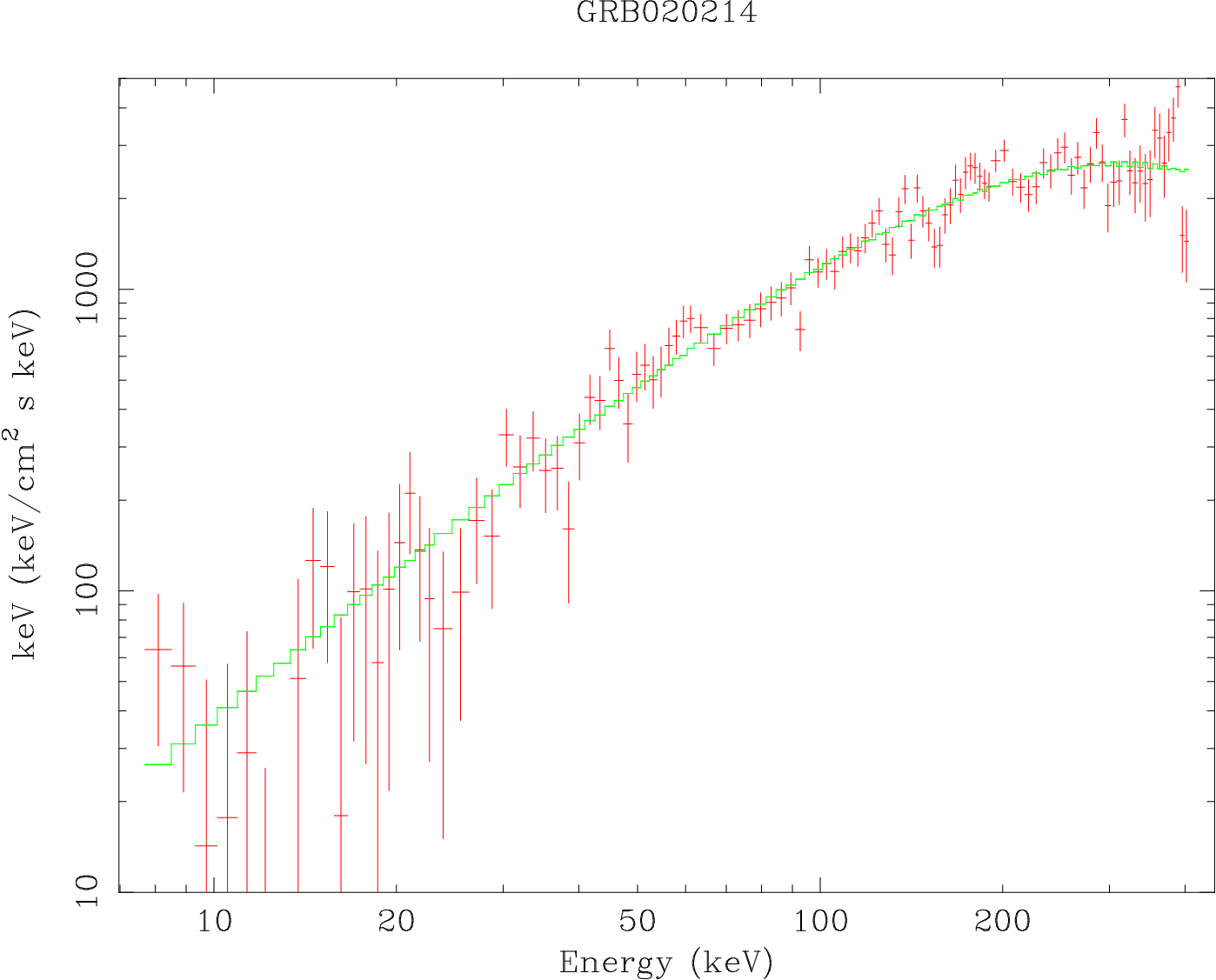} \vspace{0.5cm}

   \includegraphics[angle=-90,width=8.0cm]{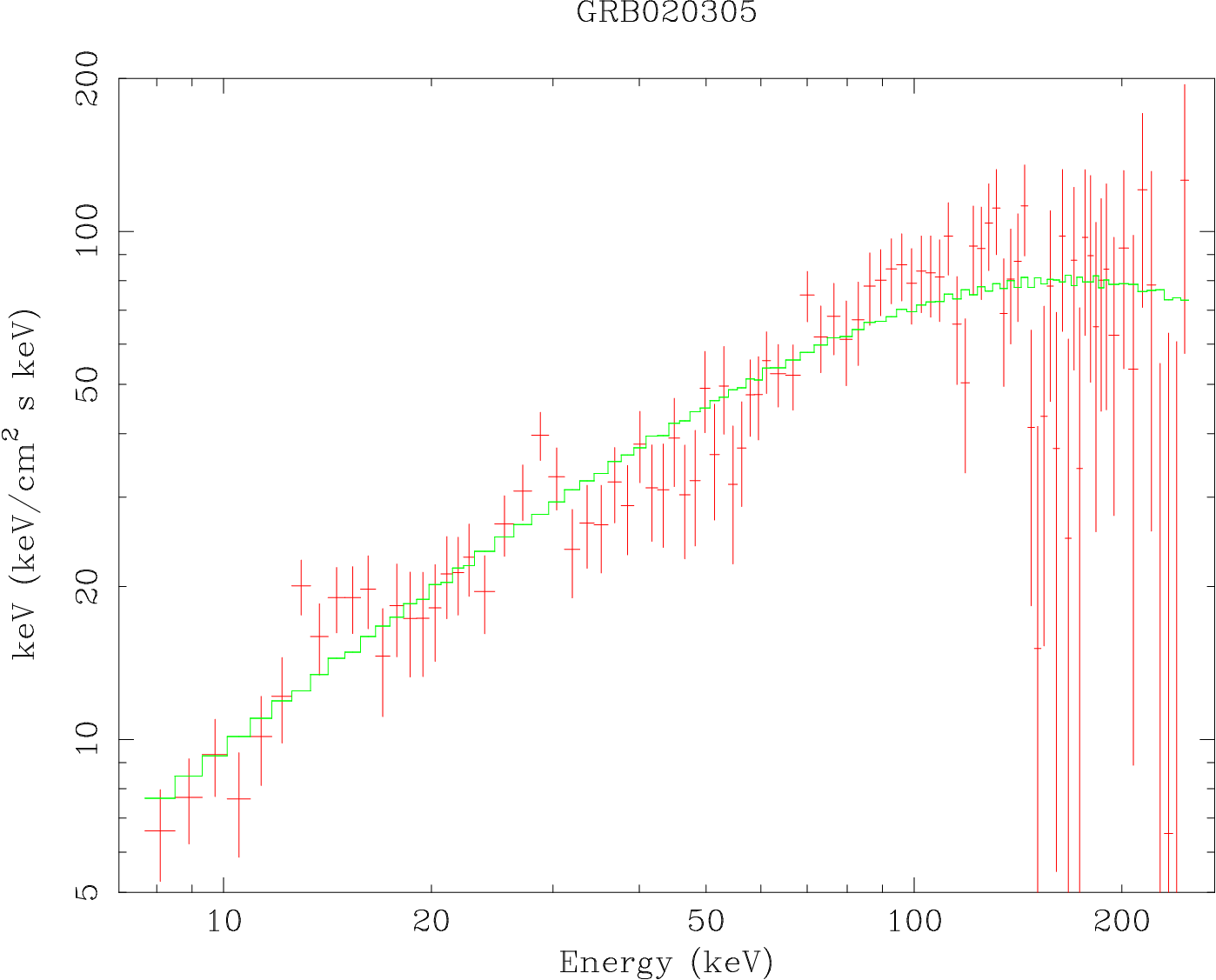} \vspace{0.5cm}

   \includegraphics[angle=-90,width=8.0cm]{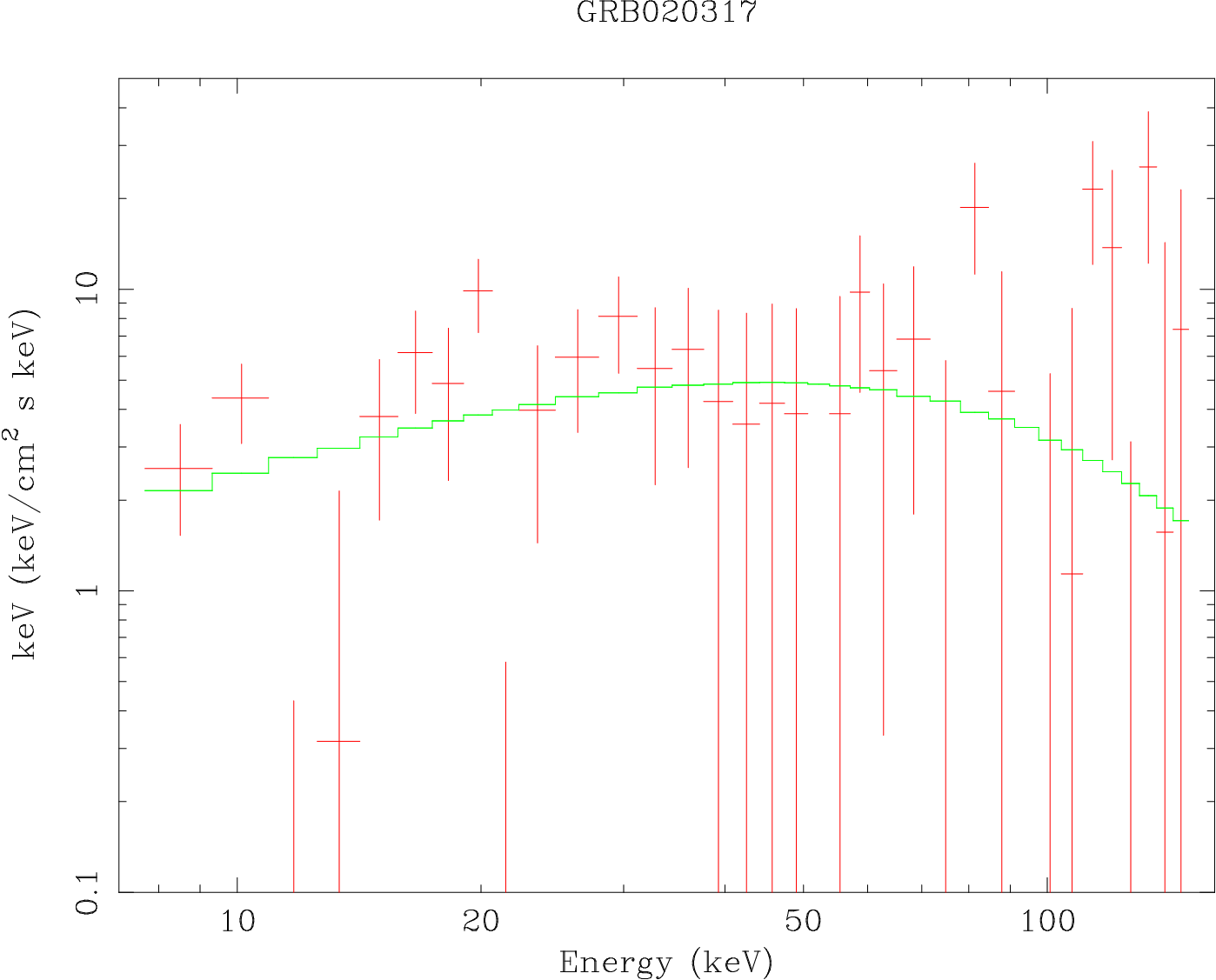}
   \end{figure}

   \begin{figure}
   \centering
   \includegraphics[angle=-90,width=8.0cm]{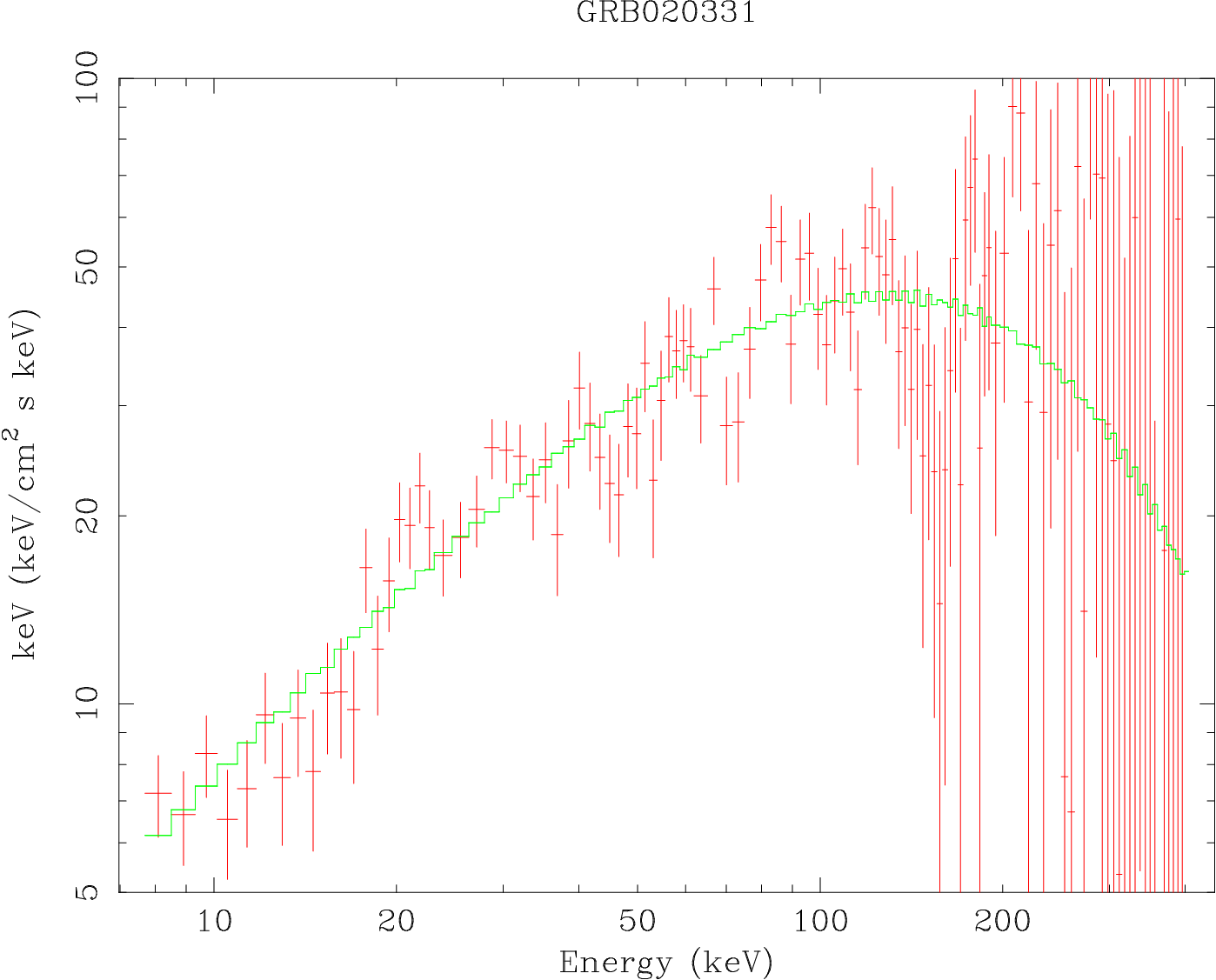} \vspace{0.5cm}

   \includegraphics[angle=-90,width=8.0cm]{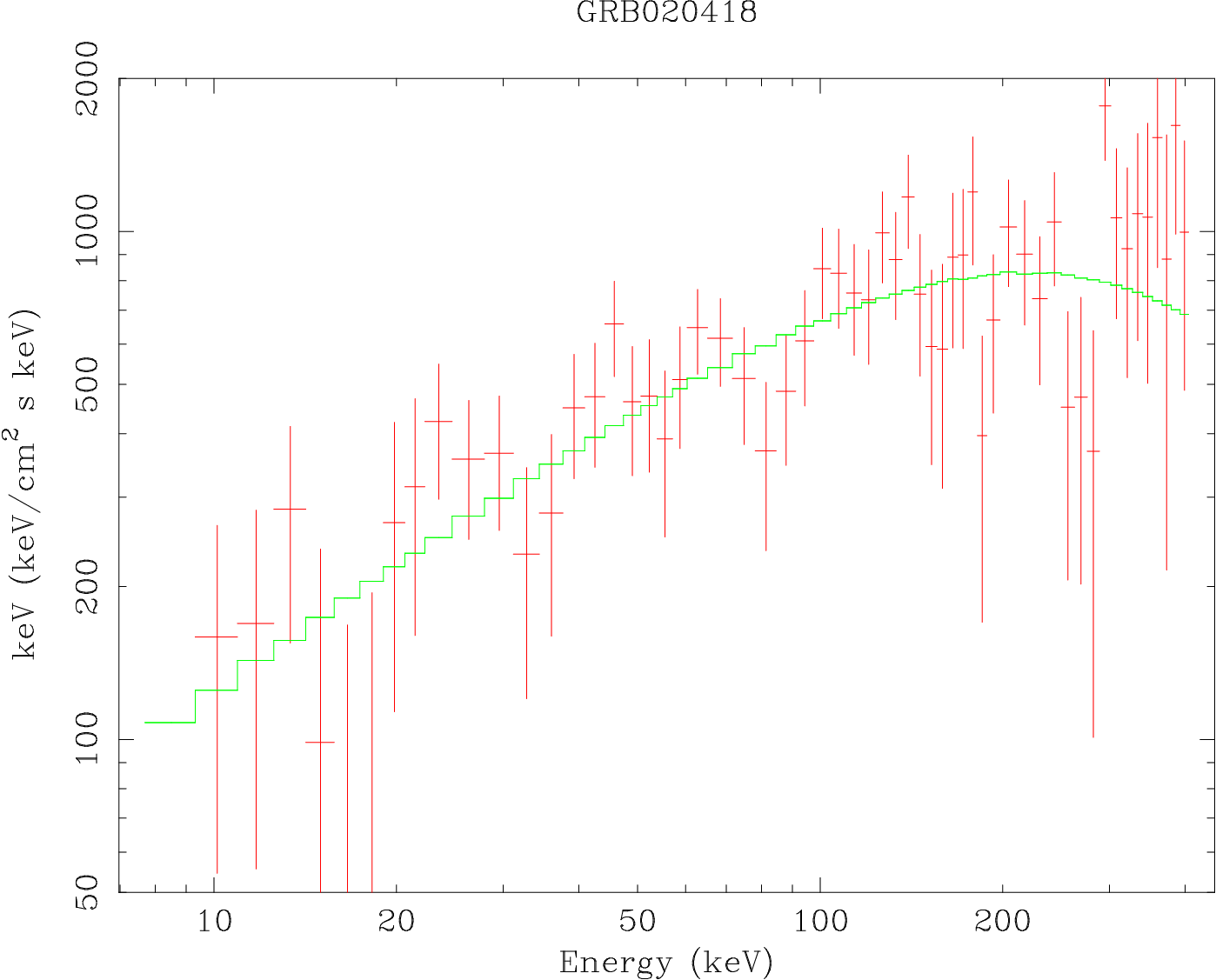} \vspace{0.5cm}

   \includegraphics[angle=-90,width=8.0cm]{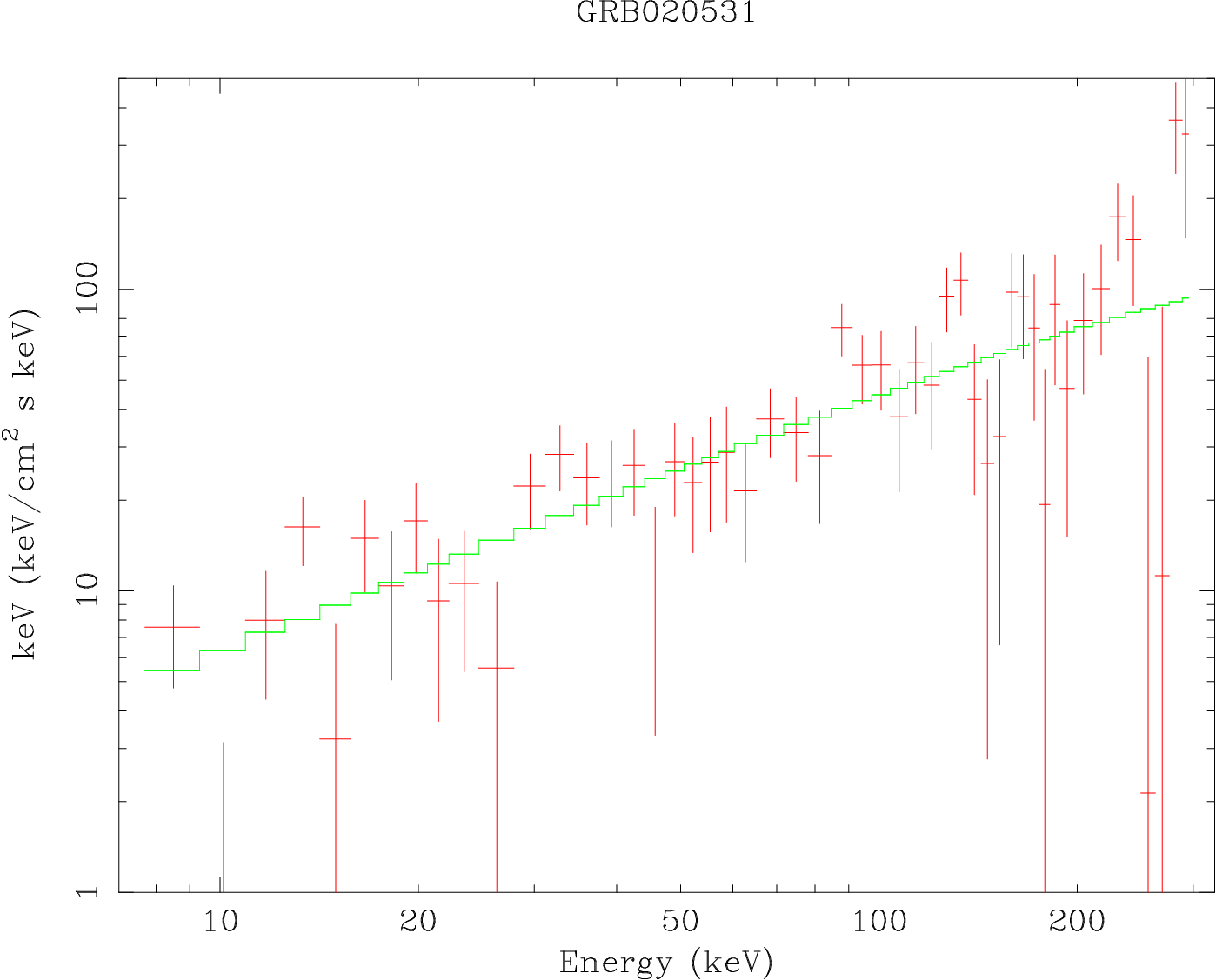} \vspace{0.5cm}

   \includegraphics[angle=-90,width=8.0cm]{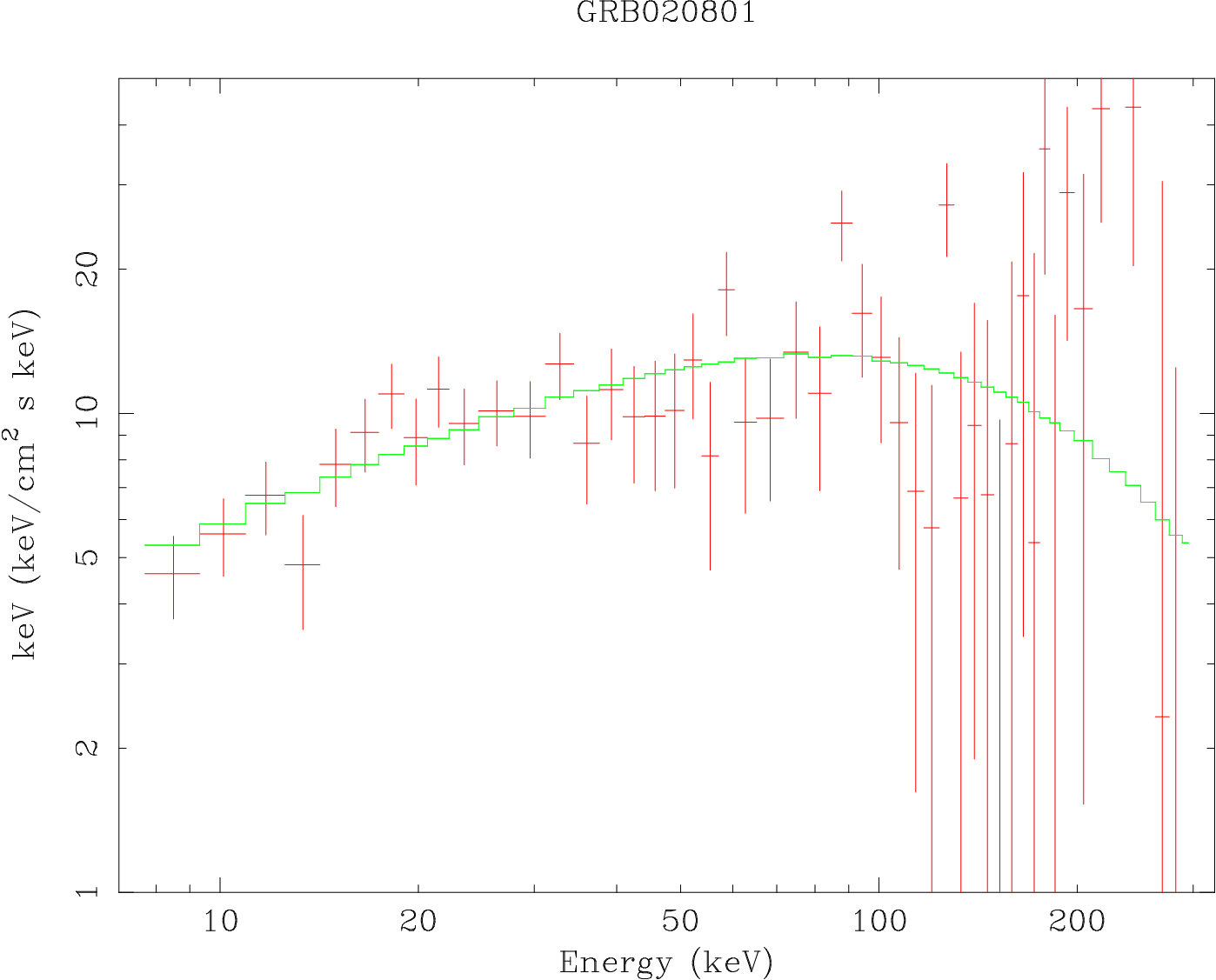} 
   \end{figure}

   \begin{figure}
   \includegraphics[angle=-90,width=8.0cm]{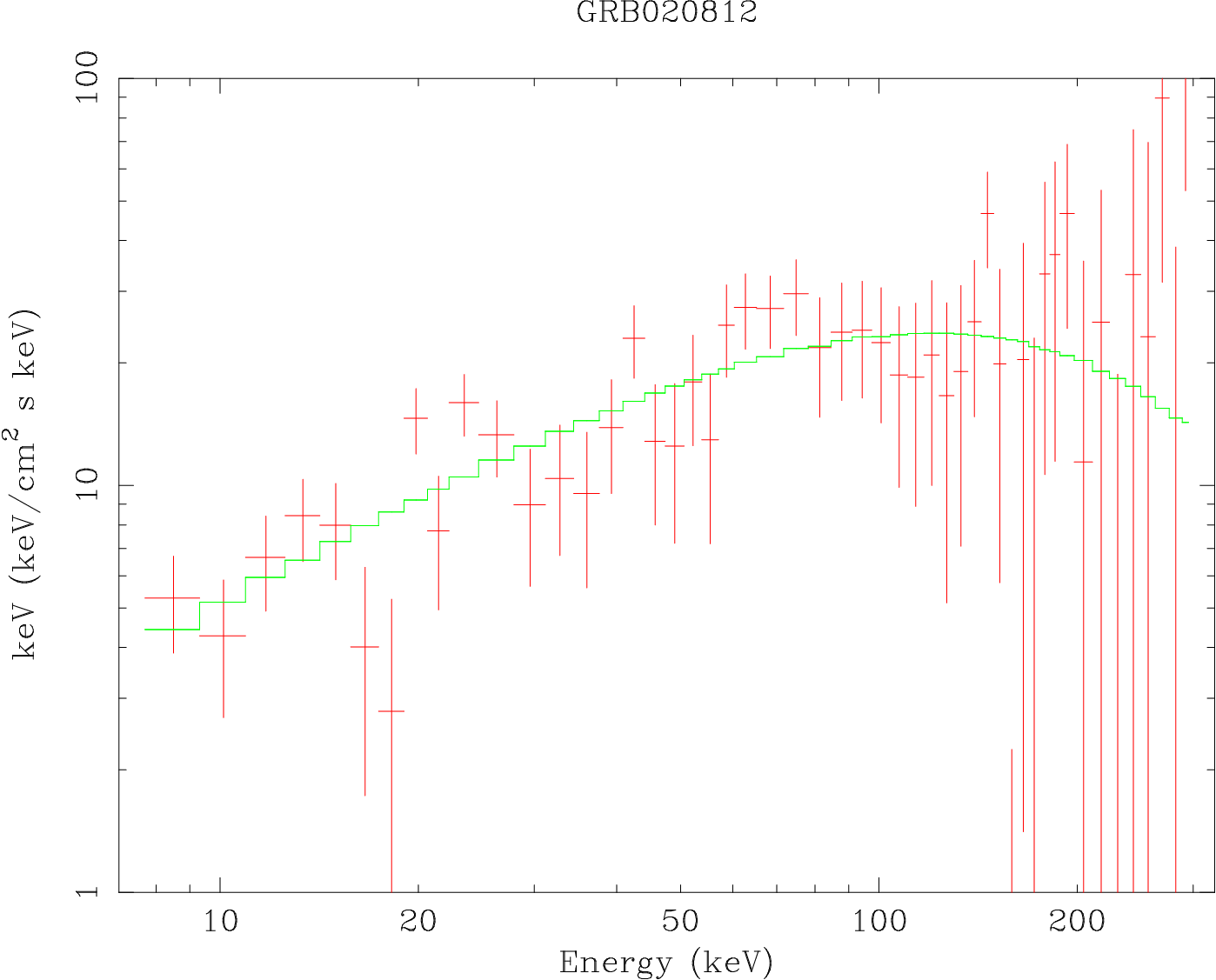} \vspace{0.5cm}

   \includegraphics[angle=-90,width=8.0cm]{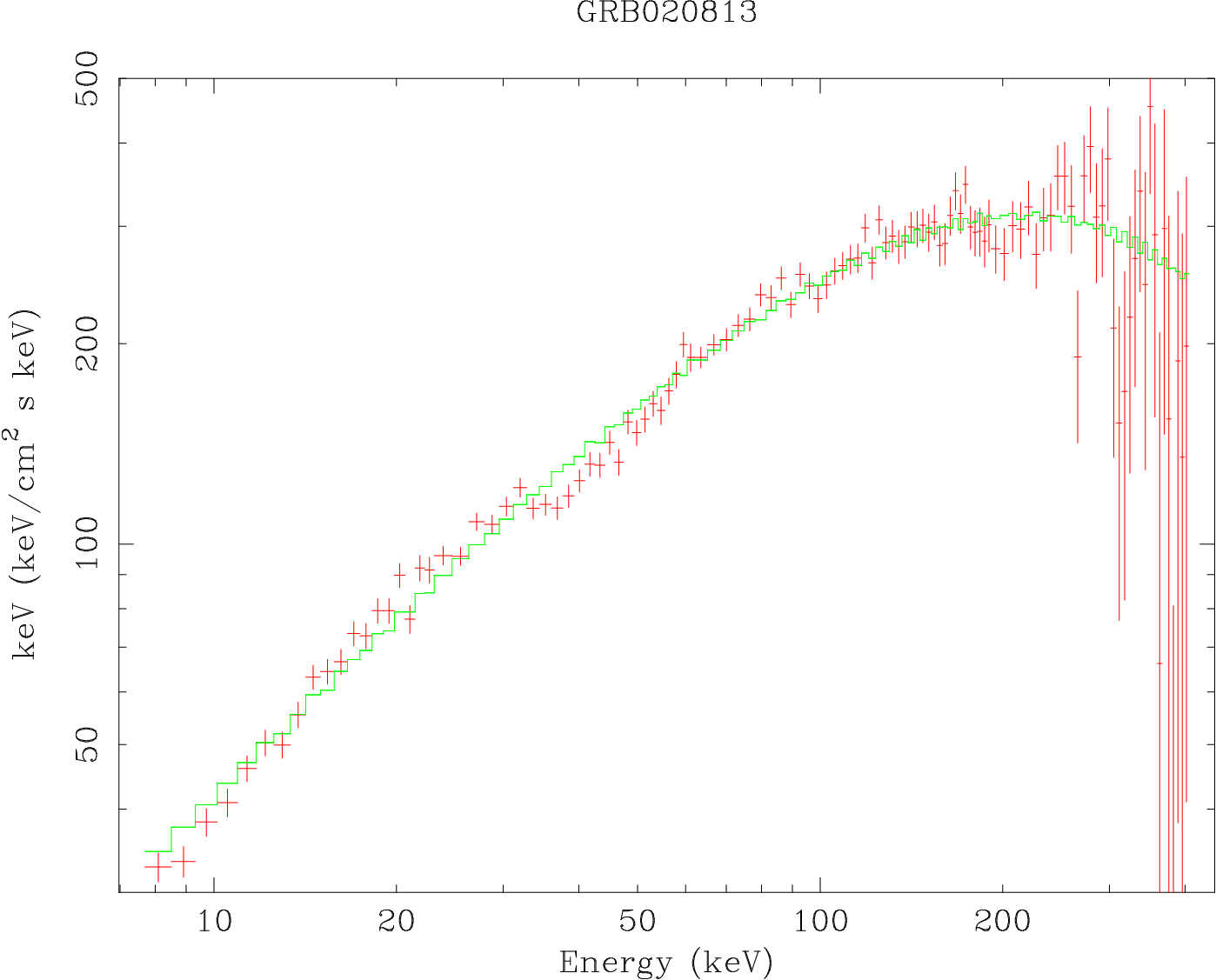} \vspace{0.5cm}

   \includegraphics[angle=-90,width=8.0cm]{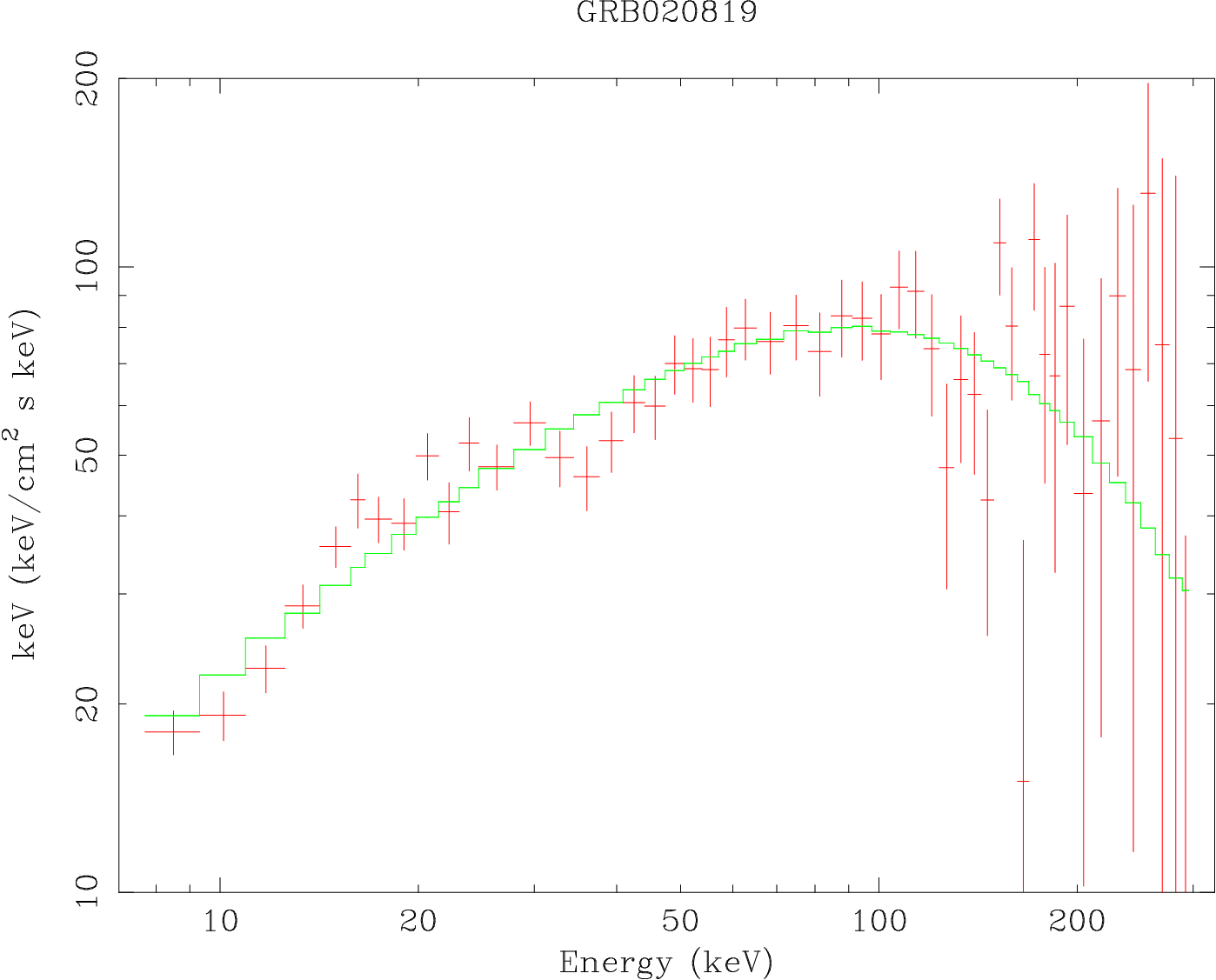} \vspace{0.5cm}

   \includegraphics[angle=-90,width=8.0cm]{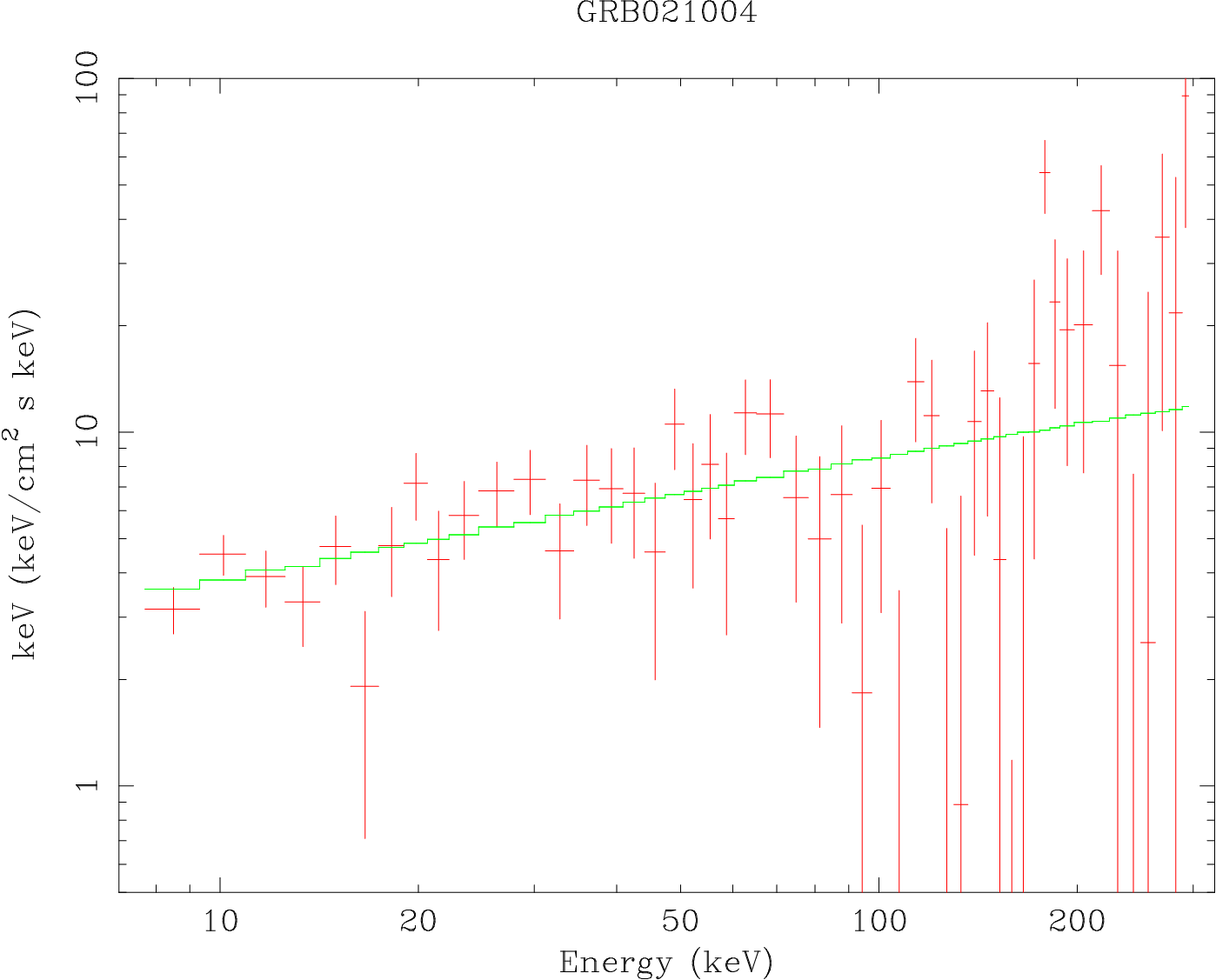} 
   \end{figure}

   \begin{figure}
   \centering
   \includegraphics[angle=-90,width=8.0cm]{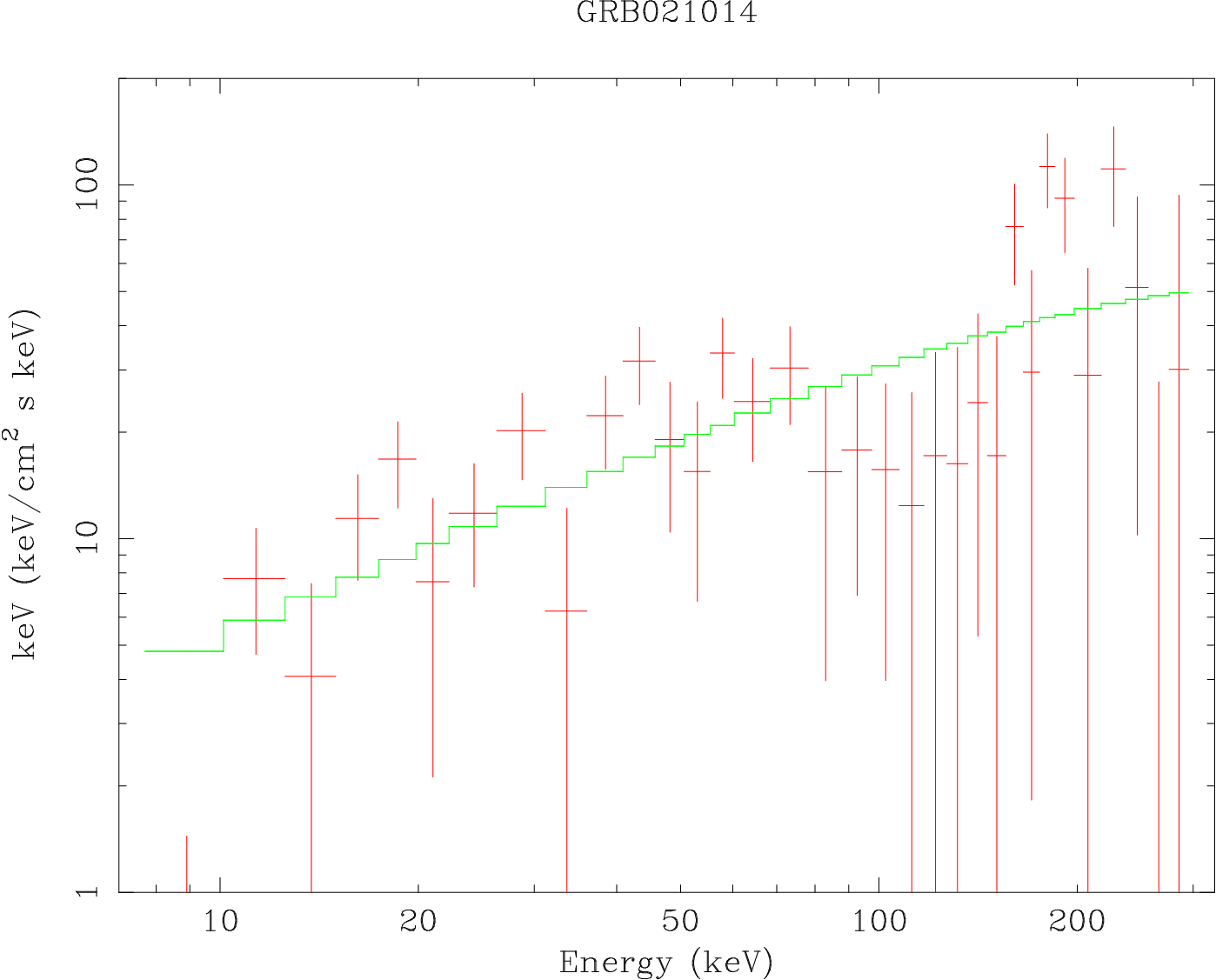} \vspace{0.5cm}

   \includegraphics[angle=-90,width=8.0cm]{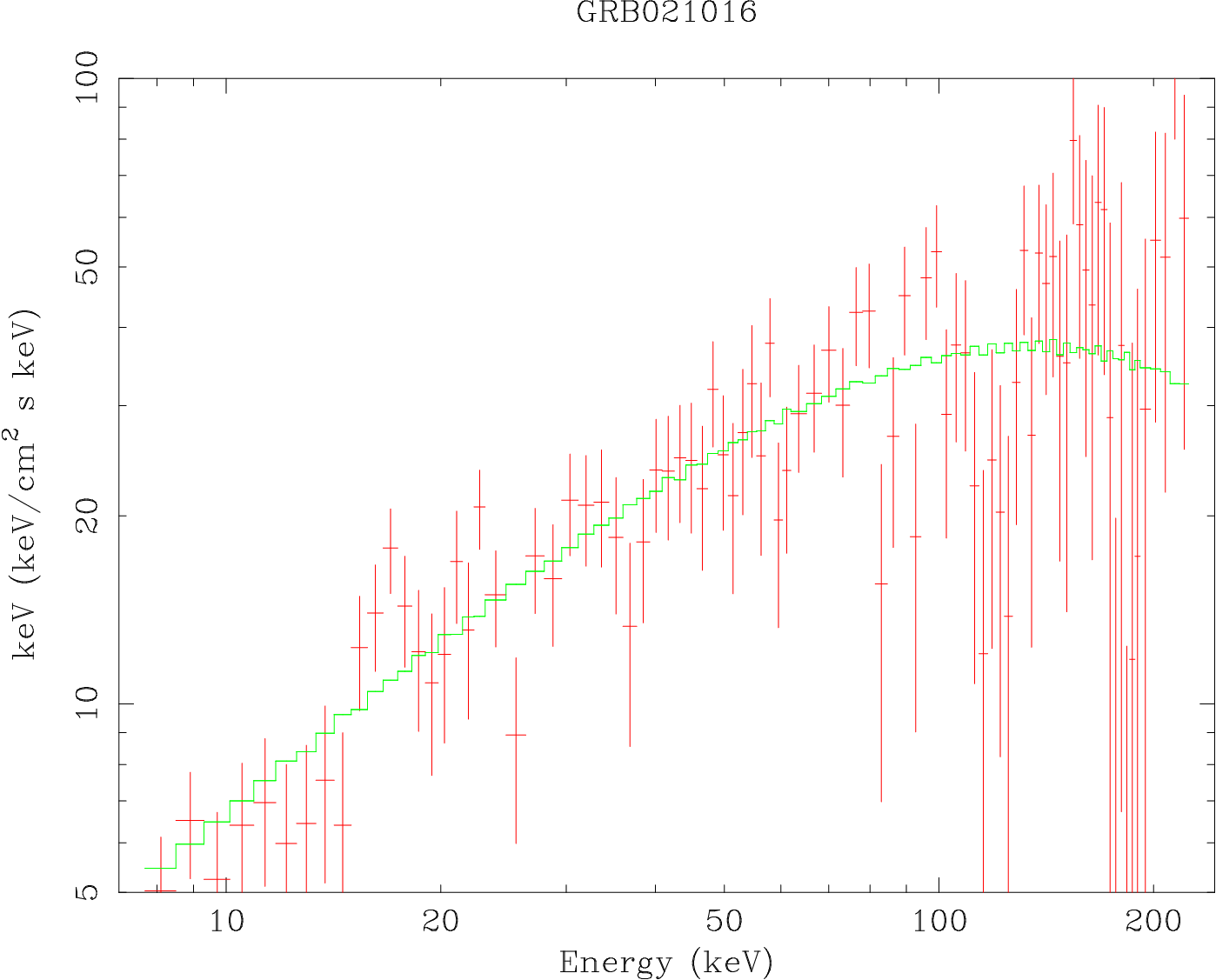} \vspace{0.5cm}

   \includegraphics[angle=-90,width=8.0cm]{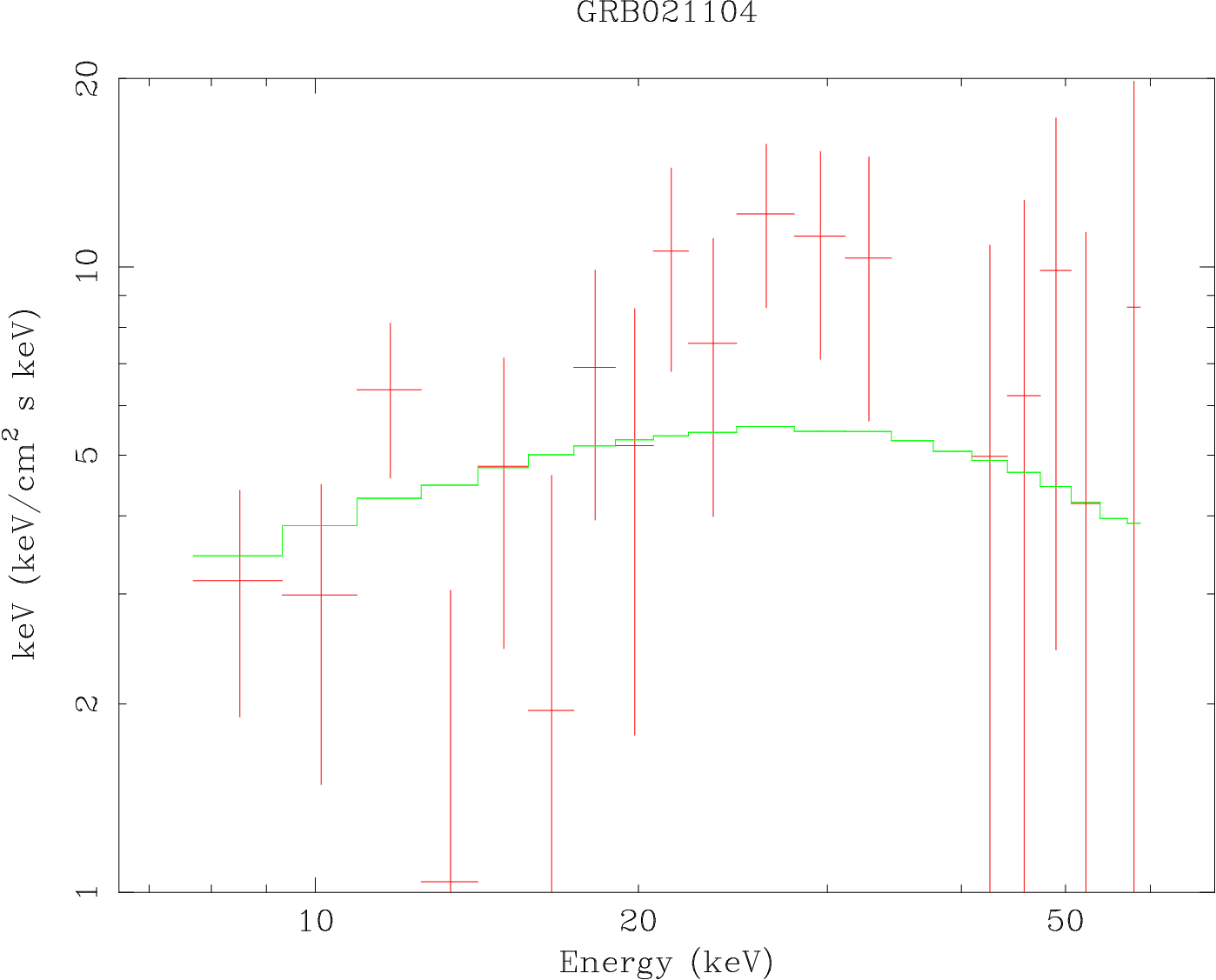} \vspace{0.5cm}

   \includegraphics[angle=-90,width=8.0cm]{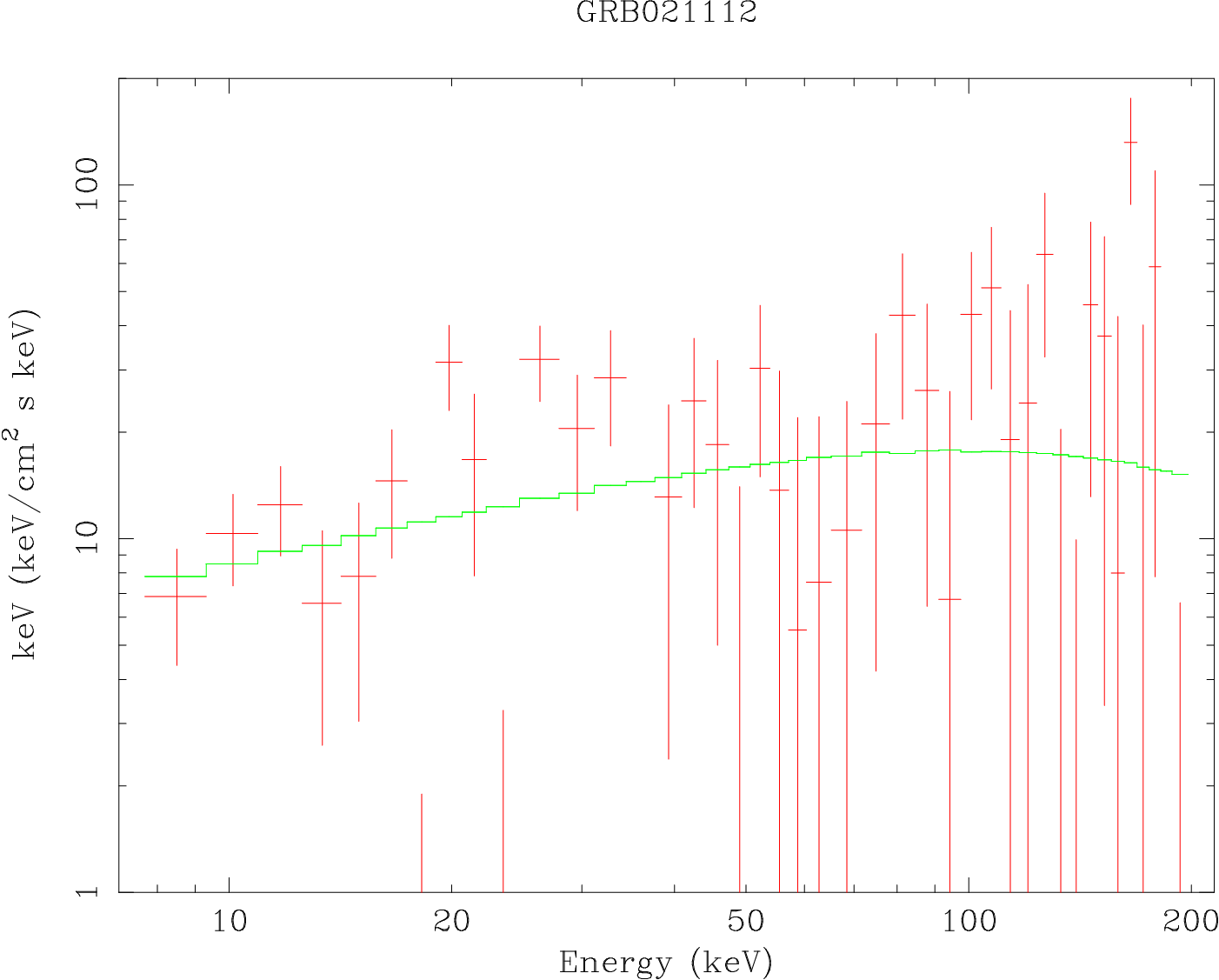} 
   \end{figure}

\end{document}